\documentclass[letterpaper]{article} 
\usepackage{arxiv}  

\usepackage{times}  
\usepackage{helvet}  
\usepackage{courier}  
\usepackage[hyphens]{url}  
\usepackage{graphicx} 
\usepackage{natbib}  
\usepackage{caption} 
\usepackage{multirow}
\usepackage{booktabs}      
\usepackage{subcaption}

\usepackage{algorithm}
\usepackage{xcolor}
\usepackage{colortbl}
\usepackage{amsmath, amssymb, amsthm}
\usepackage{cuted}

\definecolor{refpurple}{RGB}{90,60,140}
\newcommand\mcite[1]{{\color{refpurple}{\cite{#1}}}}
\newcommand\mref[1]{{\color{refpurple}{\ref{#1}}}}

\definecolor{tab1}{RGB}{255,153,153}
\definecolor{tab2}{RGB}{255,204,153}
\definecolor{tab3}{RGB}{255,246,178}

\usepackage{newfloat}
\usepackage{listings}
\DeclareCaptionStyle{ruled}{labelfont=normalfont,labelsep=colon,strut=off} 
\floatstyle{ruled}
\newfloat{listing}{tb}{lst}{}
\floatname{listing}{Listing}
\usepackage{bibentry}

\title{UMo: Unified Sparse Motion Modeling for Real-Time Co-Speech Avatars}

\author{
    Xiaoyu Zhan$^{12*}$,
    Xinyu Fu$^{1*}$,
    Chenghao Yang$^{1}$,
    Xiaohong Zhang$^{12}$,
    Dongjie Fu$^{2\dag}$, \\
    Pengcheng Fang$^{23}$,
    Tengjiao Sun$^{23}$,
    Xiaohao Cai$^{3}$,
    Hansung Kim$^{3}$,\\
    Yuanqi Li$^{1}$,
    Jie Guo$^{1}$,
    and Yanwen Guo$^{1\dag}$
}
\affiliations {
    {\Large 
        $^{1~}$Nanjing University \quad 
        $^{2~}$Mogo AI Ltd. \quad 
        $^{3~}$University of Southampton 
    }\\
    \texttt{\{zhanxy, xinyu.fu\}@smail.nju.edu.cn} \\
    \texttt{mirecofu@163.com}, \quad 
    \texttt{ywguo@nju.edu.cn}
}

\begin{document}

\makeatletter
\setlength\titlebox{5.5in} 
\def\@maketitle{
  \newcounter{eqfn}\setcounter{eqfn}{0}
  \vbox to \titlebox {
    \hsize\textwidth
    \linewidth\hsize
    \vskip 0.625in minus 0.125in
    \centering
    {\LARGE\bf \@title \par}
    \vskip 0.1in plus 0.5fil minus 0.05in
    {\Large{\textbf{\@author\ifhmode\\\fi}}}
    \vskip .2em plus 0.25fil
    {\normalsize \affiliations_\ifhmode\\\fi}
    \vskip 1em plus 2fil
    
    \begin{center}
        \centering
        \includegraphics[width=0.95\textwidth]{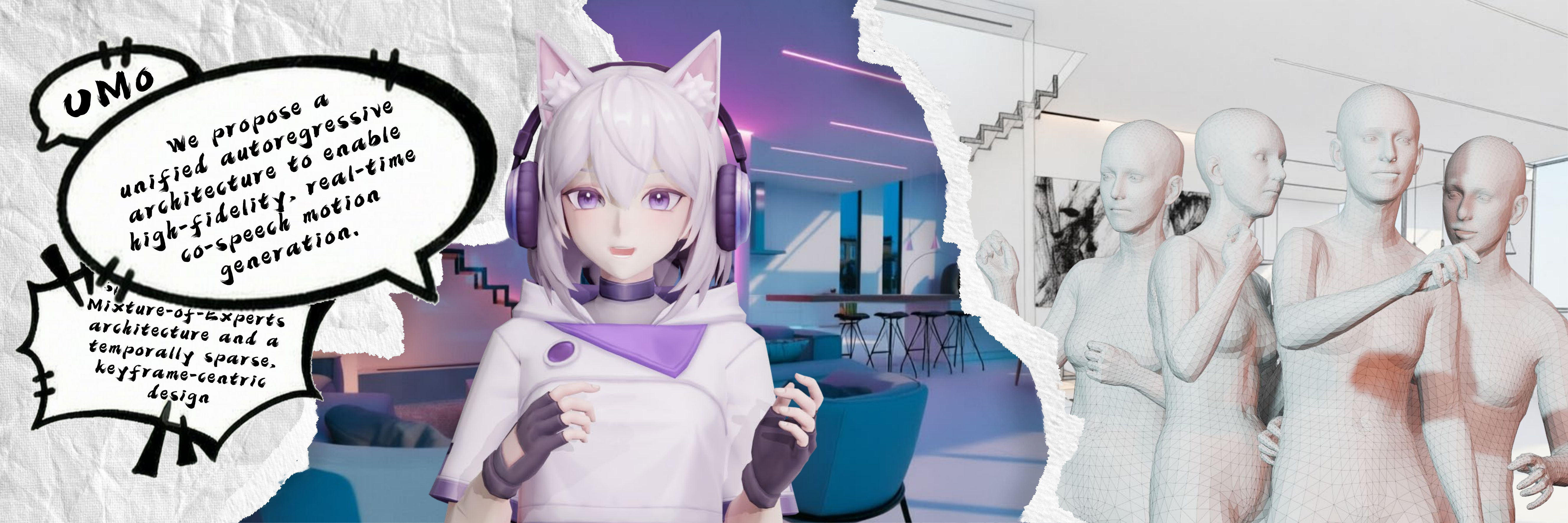}
        \captionof{figure}{\textbf{UMo} is a real-time co-speech motion generation architecture achieved through unified sparse motion modeling.}
        \label{fig:teaser}
    \end{center}
 
    \vskip 1.5em 
  }
}
\makeatother

\maketitle

\let\thefootnote\relax\footnotetext{\textsuperscript{*}Equal contribution. \textsuperscript{\dag}Corresponding author.}


\begin{abstract}
    Speech-driven gestures and facial animations are fundamental to expressive digital avatars in games, virtual production, and interactive media. However, existing methods are either limited to a single modality for audio motion alignment, failing to fully utilize the potential of massive human motion data, or are constrained by the representation ability and throughput of multimodal models, which makes it difficult to achieve high-quality motion generation or real-time performance. We present UMo, a unified sparse motion modeling architecture for real-time co-speech avatars, which processes text, audio, and motion tokens within a unified formulation. Leveraging a spatially sparse Mixture-of-Experts framework and a temporally sparse, keyframe-centric design, UMo efficiently performs real-time dense reconstruction, enabling temporally coherent and high-fidelity animation generation for both facial expressions and gestures. Furthermore, we implement a multi-stage training strategy with targeted audio augmentation to enhance acoustic diversity and semantic consistency. Consequently, UMo preserves fine-grained speech-motion alignment even under strict latency constraints. Extensive quantitative and qualitative evaluations show that UMo achieves better output quality under low latency and real-time performance constraints, offering a practical solution for high-fidelity real-time co-speech avatars.
\end{abstract}

\textbf{\section{\textbf{Introduction}}}

Speech-driven gesture and facial animation are fundamental to creating expressive, interactive digital avatars. In practical scenarios, this task is far more than a simple offline synthesis process. Instead, the system must adapt to streaming audio and changing contexts under tight time limits while maintaining natural coordination between different body parts. This requirement shifts the focus from basic motion synthesis to a more complex problem of real-time multimodal generation.

Although current methods have improved the realism of generated motion, significant obstacles remain for wide application. Most existing methods \mcite{ginosar2019learning, liu2022audio, liu2022learning, ao2022rhythmic, ao2023gesturediffuclip, zhang2024semantic, liu2025gesturelsm} rely heavily on paired audio-motion data, which are very expensive and difficult to collect at scale. This heavy dependence on limited data often results in a lack of motion diversity and poor performance in varied real-world situations. In response to this data bottleneck, emerging research \mcite{chen2024language, yu2025socialgen, hou2025motionverse, ling2024versatilemotion} has explored using Large Language Models \mcite{raffel2020exploring, touvron2023llama, bai2023qwen} to integrate diverse data sources and take advantage of their strong cross-modal capabilities.

However, existing language model based frameworks still face critical limitations in both synthesis quality and generation efficiency. In terms of quality, co-speech animation requires high-fidelity coordination between subtle facial expressions and complex body gestures, yet current methods are often restricted by their limited capacity to represent such high-dimensional human movements. More importantly, interactive avatars are exceptionally sensitive to both inference throughput and response latency. Consequently, the heavy computational cost of these models makes it very difficult to meet the stringent real-time requirements necessary for seamless, natural interaction.

In response to these problems, we propose \textbf{UMo}, a \textbf{u}nified \textbf{s}parse \textbf{mo}tion modeling architecture for real-time co-speech avatars (shown in Fig.~\mref{fig:teaser}). UMo optimizes the trade-off between quality and efficiency by introducing spatial sparsity and temporal sparsity into its architecture. It also integrates text, audio, and motion tokens into a single autoregressive formulation, enabling versatile training across multimodal tasks and diverse data sources while maintaining the high throughput required for real-time inference.

Spatial sparsity is achieved via a Mixture-of-Experts (MoE) framework to overcome the representational limitations of language models in action generation. By routing distinct body regions to dedicated expert networks, this framework significantly enhances the fidelity and coherence of generated motions while maintaining high throughput for streaming inference. Simultaneously, temporal sparsity is introduced through our keyframe-centric design. Considering the high redundancy in adjacent frames of motion sequences, we reformulate full-sequence prediction into a two-stage paradigm, where the language model predicts only sparse but critical keyframes. To preserve continuous details during this sparse generation process, we incorporate a light-weight interpolation network that reconstructs non-keyframe intervals conditioned on fixed keyframe anchors. Collectively, these mechanisms enable real-time, dense reconstruction of synchronized facial expressions and gestures with both high fidelity and low latency. 

During training, we employ a three-stage strategy to stabilize the optimization of multimodal objectives. The training process starts with foundational motion modeling, subsequently proceeds to fine-grained audio-motion alignment, and finally in end-to-end autoregressive synchronization. We further incorporate a targeted audio augmentation strategy to mitigate paired data scarcity. Extensive experiments demonstrate that our method achieves high-quality co-speech generation while maintaining real-time performance within a large language model framework.

In summary, our contributions are fourfold.

\begin{itemize}
\item We propose a unified autoregressive architecture that jointly models text, audio, and motion tokens to enable high-fidelity, real-time co-speech motion generation.
\item We develop a spatially sparse MoE backbone that scales model capacity and representational power without increasing the per-token computational overhead during streaming inference.
\item We introduce a temporally sparse keyframe framework combined with interpolation for dense reconstruction, achieving real-time performance while preserving fine-grained motion details.
\item We design a practical training pipeline, including a multi-stage training recipe and a targeted audio augmentation strategy, to enhance the generation quality and robustness in co-speech scenarios.
\end{itemize}

\textbf{\section{Related Work}\label{sec:related_work}}

\textbf{\subsection{Text-to-Motion Generation}}

Text-to-motion synthesis focuses on generating high-fidelity human motion sequences conditioned on natural language descriptions. The rapid evolution of this field has been fundamentally underpinned by several large-scale annotated motion datasets \mcite{mahmood2019amass, guo2022humanml3d, punnakkal2021babel, lin2023motionx, rempe2026kimodo}, which provide the necessary supervision for cross-modal alignment.

Early text-to-motion research focuses on establishing foundational shared representations between text and motion \mcite{petrovich2021temos, guo2022tm2t, tevet2022motionclip}. Subsequent methods have been dominated by continuous diffusion frameworks \mcite{zhang2022motiondiffuse, tevet2023human, yuan2023physdiff, zhang2023remodiffuse, chen2024executing, zhou2024emdm, zhang2023finemogen, zhang2025energymogen, karunratanakul2023guided, xie2024omnicontrol, zhang2025motionduet} and discrete sequence modeling \mcite{zhang2023generating, guo2024momask, pinyoanuntapong2024mmm, chen2024language,jiang2024motiongpt, wu2024motionagent, zeng2025light, zhu2025motiongpt3, zhang2024motiongptfinetuned, lu2025scamo, fu2026mogo, li2026llamo}. More recently, some research also attempts to integrate these two paradigms \mcite{chi2024m2d2m, zhao2025dartcontrol, xiao2025motionstreamer, nazarenus2026actionplan, zhang2026dimo, gu2026bridging}. While continuous diffusion frameworks achieve superior motion fidelity and fine-grained controllability, they are often hindered by the computational overhead of iterative sampling. Conversely, discrete generative paradigms prioritize causal and streaming generation, seeking to optimize the balance between inference efficiency and synthesis quality. 

Another growing trend is unified modeling, where motion is represented as tokens or latents co-trained with text and other modalities for joint generation and understanding \mcite{chen2024language, jiang2024motiongpt, wu2024motionagent, zhang2024motiongptfinetuned, li2026llamo}. Researchers aim to employ a unified and streamlined architecture to explore scaling laws in the field of motion generation, leveraging massive amounts of data. Along similar lines, UMo embeds text, audio, and motion within a jointly learned discrete space. This unified representation, coupled with an autoregressive modeling paradigm, allows for the seamless streaming motion generation over arbitrary temporal scales.

\textbf{\subsection{Audio-to-Motion Synthesis}}

Additionally, some studies have explored using audio as the guidance signal, covering distinct sub-tasks such as music-driven dance generation \mcite{alexanderson2023listen} and co-speech motion synthesis \mcite{ginosar2019learning, liu2022learning, liu2022audio, ao2022rhythmic, ao2023gesturediffuclip, xing2023codetalker, sun2024diffposetalk, chen2024diffsheg, zhao2024media2face, liu2025gesturelsm, cai2025mio}, which are supported by multi-modal datasets featuring synchronized audio and motion streams \mcite{li2021ai, liu2022beat, yi2023generating, liu2024emage}. Although both fields leverage audio rhythm to guide human motion, their underlying objectives diverge significantly. Specifically, music-driven models prioritize long-horizon temporal consistency and fluency, whereas co-speech frameworks demand fine-grained alignment between speech and body gestures, as well as facial expressions and lip movements. Furthermore, robust co-speech systems should also effectively model naturalistic idle-state transitions during silent intervals.

Our work specifically addresses co-speech synthesis. Early research primarily focuses on the generation of body motion and gestures \mcite{ginosar2019learning, liu2022audio, liu2022learning, ao2022rhythmic, ao2023gesturediffuclip, zhang2024semantic}. These studies largely attempt to align acoustic features with motion representations while enhancing conditional control capabilities. Conversely, a separate branch concentrates on the synchronization of facial expressions and lips with speech input \mcite{xing2023codetalker, sun2024diffposetalk, zhao2024media2face, pan2025model, han2025tiny}. However, the synthesis of isolated parts faces significant constraints in real-world scenarios, where head and body movements are intrinsically coupled. 

To address this limitation, recent advancements have shifted toward the holistic generation of bodies, gestures, and expressions within unified frameworks \mcite{yi2023generating, liu2024emage, chen2024diffsheg, xu2024mambatalk, liu2025gesturelsm, cai2025mio}. This emerging paradigm investigates diverse architectural strategies including end-to-end, cascaded, and parallel configurations to achieve seamless cross-modal synthesis. Specifically, UMo utilizes a unified end-to-end framework to encode audio, motion, and expressions. Considering the distinct properties of body movements, we categorize them into upper body, lower body, and gestures. Combined with facial expressions, these modalities are processed via four expert networks. For each token generation, UMo dynamically selects and activates one experts to optimize the trade-off between computational cost and generation quality.

\textbf{\subsection{Real-Time Co-Speech Avatars}}

Driven by the needs of real-world applications, researchers have consistently pursued the ability of models to respond in real-time. This imperative requires network pipelines to support causal or streaming motion generation. Therefore, while full-sequence bidirectional diffusion models \mcite{chen2024diffsheg} excel at generating high-quality motion sequences, their inherent reliance on full-context attention makes them ill-suited for real-time generation. In recent years, some efforts to accelerate diffusion, such as latent diffusion \mcite{zhao2025dartcontrol}, few-step distillation \mcite{chern2025livetalk}, and streaming denoising \mcite{nazarenus2026actionplan, cai2025mio}, have provided new possibilities for diffusion based real-time generation. For more practical considerations, most work \mcite{tevet2024closd, chen2024taming, zhao2025dartcontrol, xiao2025motionstreamer, zhang2025primal} is being attempted to combine autoregressive to achieve real-time motion generation. Because it naturally adapts to the streaming generation. However, due to the conflict between generation speed and the window size of causal attention, these methods are prone to losing their perception of historical sequence information.

Within the co-speech domain, the synthesis task has grown increasingly complex with multi-modal inputs and diverse output spaces. Regarding audio representation, MIBURI \mcite{mughal2026miburi} leverages the full-duplex speech-text foundation model Moshi \mcite{defossez2024moshi}, whereas others approaches \mcite{xu2024mambatalk, liu2025gesturelsm} typically employ separate audio encoders to extract acoustic features. Distinctly, MIO \mcite{cai2025mio} adopts a distinct strategy by introducing a dedicated Thinker module designed to bridge the gap between input modalities and downstream generative components. 

In terms of motion representation, most studies \mcite{mughal2026miburi, xu2024mambatalk, liu2025gesturelsm} employ a decoupled encoding paradigm that partitions the human body into functional regions including the face, hands, and upper and lower body. This modular design allows the system to combine parts during inference, which reduces the need for massive datasets and simplifies the processing of full-body movements. MIO \mcite{cai2025mio} adopts a more aggressive strategy by utilizing independent body and face animators for parallel motion synthesis. To address potential alignment challenges arising from such decoupled representations, GestureLSM \mcite{liu2025gesturelsm} introduces the spatial-temporal modeling to explicitly capture the interactions across different body regions. 

Regarding inference efficiency, MIBURI \mcite{mughal2026miburi} integrates motion embeddings into a lightweight foundation model to facilitate efficient autoregressive generation. Alternatively, GestureLSM \mcite{liu2025gesturelsm} utilizes shortcut flow-matching to achieve real-time performance through efficient sampling.

To address the limitations of fixed window lengths and generation speed, UMo employs keyframes to significantly accelerate generation and extend the effective length of historical information. Our approach simultaneously minimizes temporal redundancy in adjacent frames and boosts inference efficiency without introducing additional computational overhead.

\textbf{\section{Overview}\label{sec:overview}}

\begin{figure*}[t]
  \includegraphics[width=0.97\textwidth]{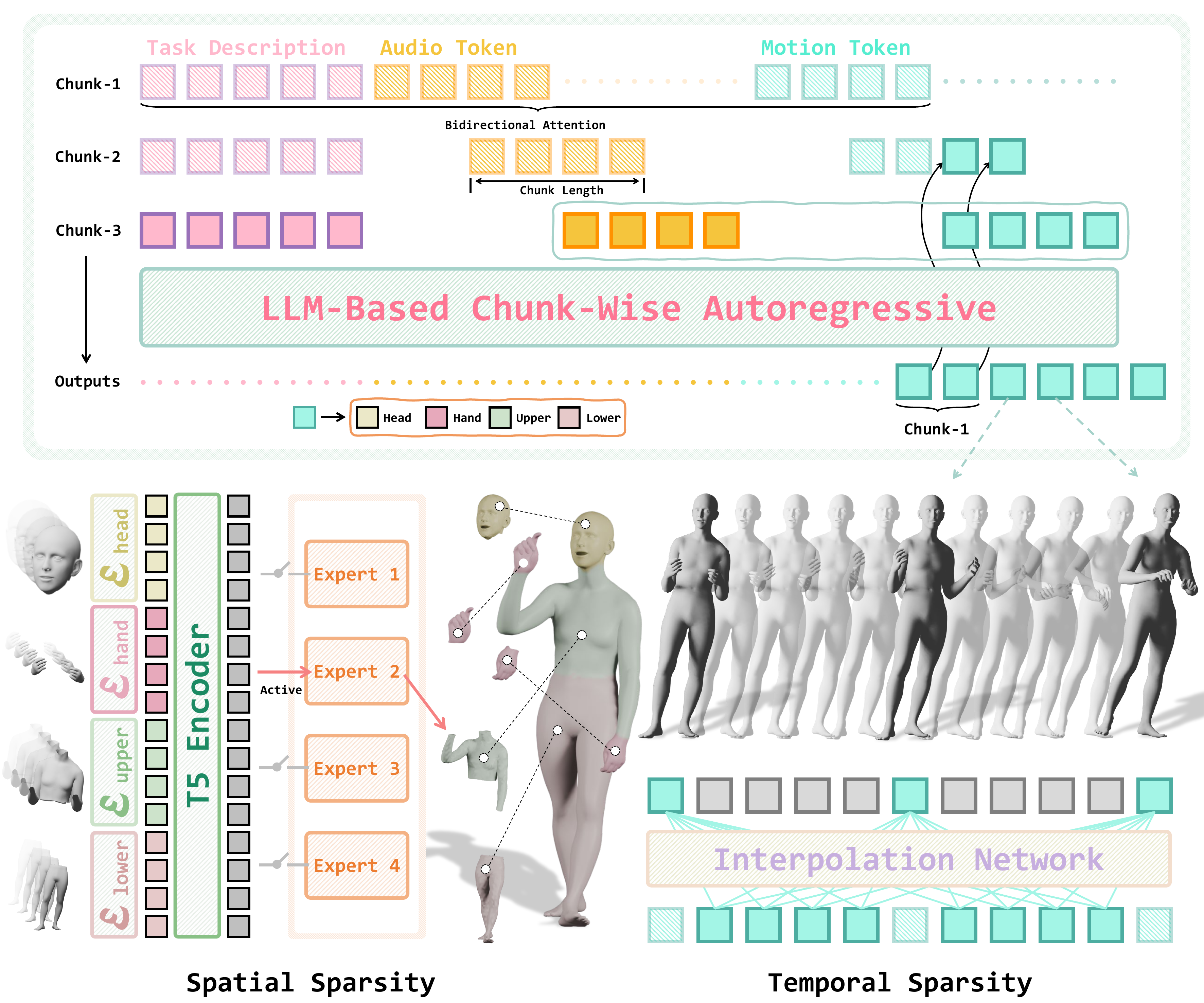}
  \caption{\textbf{Pipeline overview}. UMo is a unified sparse architecture for real-time co-speech generation. We achieve both high-fidelity motion generation and low-latency response using an LLM-based backbone through an autoregressive framework (Sec.~\mref{sec:autoregressive_design}), combined with spatial sparsity (Sec.~\mref{sec:method_moe}) and temporal sparsity (Sec.~\mref{sec:method_interp}).}
  \label{fig:pipeline}
\end{figure*}

In this work, we formally define \textbf{real-time co-speech generation} as follows.
Let $\mathbf{a}_{1:t}$ denote observed speech signals up to time $t$ (audio with optional text/task cues), and let
\begin{equation}
\mathbf{m}_{1:t}
=
\big(\mathbf{g}_{1:t}, \mathbf{e}_{1:t}\big)
\end{equation}
denote synchronized full-body co-speech motion, where $\mathbf{g}$ represents gestures/body dynamics and $\mathbf{e}$ represents facial expressions.
The goal is to learn a causal policy that predicts future motion from past observations only under strict online latency constraints:
\begin{equation}
p_\theta(\mathbf{m}_{t+1}\mid \mathbf{a}_{1:t},\mathbf{m}_{1:t}).
\end{equation}
Here, $p_\theta(\cdot)$ denotes the model conditional distribution parameterized by learnable parameters $\theta$.

In practice, due to the particularity of the co-speech task itself, we should explicitly account for how audio after time $t$ is obtained.
Under the same notation, we consider two common regimes.

\noindent\textbf{Offline audio available.} When target-horizon audio is precomputed or already observed offline, motion is generated with look-ahead audio:
\begin{equation}
p_\theta\!\left(\mathbf{m}_{t+1}\mid \mathbf{a}_{1:t+1},\mathbf{m}_{1:t}\right).
\end{equation}

\noindent\textbf{Joint motion-audio co-generation.} When future audio is not given, the model jointly generates synchronized future audio and motion:
\begin{equation}
p_\theta\!\left(\mathbf{m}_{t+1},\mathbf{a}_{t+1}\mid \mathbf{a}_{1:t},\mathbf{m}_{1:t},\mathbf{c}\right),
\end{equation}
where $\mathbf{c}$ denotes exogenous control conditions for audio generation (e.g., user queries, dialogue intent, or environment changes).
This can also be factorized autoregressively as
\begin{equation}
p_\theta(\mathbf{a}_{t+1}\mid \mathbf{a}_{1:t},\mathbf{m}_{1:t},\mathbf{c})\cdot
p_\theta(\mathbf{m}_{t+1}\mid \mathbf{a}_{1:t+1},\mathbf{m}_{1:t},\mathbf{c}).
\end{equation}

In our implementation, we instantiate $\mathbf{m}$ with compositional motion token streams. We assume that the streaming audio is known, UMo autoregressively predicts synchronized tokens for four parts (face, hands, upper body, and lower body), denoted by
$\mathbf{z}^{f}_{1:t}$, $\mathbf{z}^{h}_{1:t}$, $\mathbf{z}^{u}_{1:t}$, and $\mathbf{z}^{l}_{1:t}$.
This compositional layout follows LOM settings \mcite{chen2024language}.

Audio is encoded by a HuBERT feature extractor \mcite{hsu2021hubert} with k-means quantization (L9, $K_a{=}500$), yielding one discrete audio token every $320$ waveform samples.
At $16$kHz this corresponds to $50$ audio tokens per second.
Motion is encoded by four region-specific VQ tokenizers (inherited from LOM\mcite{chen2024language}) with codebook size $K_f{=}K_h{=}K_u{=}K_l{=}256$ and a frequency of $30$ tokens per second. 
At the token level, we have
\begin{equation}
\mathbf{a}_{1:t} \in \{1,\dots,K_a\}^{t}, \quad
\mathbf{z}^{p}_{1:t} \in \{1,\dots,K_p\}^{t},\ p\in\{f,h,u,l\}.
\end{equation}
During decoding, each predicted stream $\mathbf{z}^{p}_{1:T}$ is mapped back by its corresponding VQ decoder to continuous parameters for that body region.

\textbf{\section{Method}\label{sec:method}}

We introduce \textbf{UMo}, a unified sparse autoregressive architecture for real-time co-speech generation. As shown in Fig.~\mref{fig:pipeline}, UMo serializes audio, task cues, and motion history into tokens. Leveraging a language model, UMo integrates diverse tasks into a unified training framework, utilizing multi-modal data to jointly optimize and enhance co-speech generation capabilities. Its general language processing capability also helps improve the model's expressiveness and speech generalization ability Furthermore, we achieve high quality and efficiency by combining spatial sparsity, implemented through a routed MoE backbone, with temporal sparsity via keyframe-centric prediction. These designs enable high-fidelity co-speech animation under strict latency and bounded compute constraints.

This section presents the implementation details of our architecture, training process, and the data modifications employed for UMo. Including chunk-wise autoregressive (Sec.~\mref{sec:autoregressive_design}); spatially sparse MoE framework (Sec.~\mref{sec:method_moe}); temporally sparse keyframe framework (Sec.~\mref{sec:method_interp}); training recipe and audio augmentation (Secs.~\mref{sec:method_training}).

\textbf{\subsection{Chunk-Wise Autoregressive}
\label{sec:autoregressive_design}}

In the context of real-time co-speech tasks, user experience is determined not only by motion quality but also by perceived latency. Previous real-time works based on diffusion usually use full-sequence motion generation \cite{liu2025gesturelsm}. However, due to the substantial computational overhead of Large Language Models and their lower throughput efficiency, achieving real-time performance in sequence generation is infeasible. In contrast, the autoregressive framework significantly accelerates single-step inference by reducing the action window size during inference. This capability provides users with immediate motion feedback, mitigates long response latencies, and ultimately enhances the sense of immersion.

Apart from this, under strict real-time inference budgets, retaining the full context history is often computationally impractical.
To bound latency and memory while preserving recent temporal dependencies, we approximate history conditioning with a fixed-size sliding window. Such chunk-wise autoregressive framework is tailored specifically for real-time interactive co-speech avatars.

While aiming to balance inference speed and generation quality, we observed that purely causal architectures offered limited benefits. Therefore, we retained the lightweight Flan-T5-Base backbone \mcite{raffel2020exploring} from LOM \mcite{chen2024language}, enhancing the encoding stage with additional contextual information to improve generation quality. Unlike fully causal autoregressive models, a key advantage of T5 as a prefix language model is its hybrid attention mechanism. It applies bidirectional attention to the historical prefix while maintaining causal attention for decoding. This enables each historical token to attend to the complete input context, ensuring a robust comprehension of historical data. This capability is vital for our complex multimodal framework, as it ensures that different modalities can fully perceive and interact with each other.

Concretely, for each target prediction window in our implementation, we introduce a history sliding-window.
Let
\begin{equation}
\mathbf{m}_{t:t+L-1}
=
\big[\mathbf{m}_{t:t+P-1},\ \mathbf{m}_{t+P:t+L-1}\big],
\end{equation}
and let $\mathbf{a}_{t:t+L-1}$ denote the aligned speech condition window.
The implemented objective is
\begin{equation}
p_\theta\!\left(\mathbf{m}_{t+P:t+L-1}\mid \mathbf{m}_{t:t+P-1},\mathbf{a}_{t:t+L-1}\right).
\end{equation}
The model leverages a motion prefix and synchronized audio to forecast future movements. To maintain temporal consistency, the prefix is locked as immutable history while only the future segment is generated.

The input data consists of three primary components. First, a textual task description is provided. Second, the audio input is rate-aligned with the motion and segmented by window size to cover both the full prefix history and the target window during a single inference step. Finally, the motion inputs are processed, where prefix motions are divided into four regions covering the face, hands, upper body, and lower body, with continuous encoding applied to each part. During the encoding phase, UMo employs a bidirectional attention mechanism to globally model all these inputs.

Specifically, for each chunk, we keep $P$ frames previous motion tokens as prefix and predict future $N$ frames tokens for all four parts. Formally, for chunk index $c$, we build
\begin{equation}
\hat{\mathbf{z}}^{*}_{c}
=
P_{\theta}\!\left(
a_{t_c-P:t_c+N},\;
\mathbf{z}^{*}_{t_c-P:t_c-1}
\right),
\quad * \in \{f,h,u,l\},
\end{equation}
where $P_{\theta}$ is the conditional predictor. After prediction, each token stream is decoded by its corresponding part decoder. The choice of chunk size significantly influences both the quality and efficiency of inference. Notably, the impact of chunk size on the first-frame delay and inference speed is not linear. Specifically, as the chunks become smaller, the marginal gains in speed diminish. Consequently, we set $P=10$ and $N=5$ for our final model.

\textbf{\subsection{Spatial Sparse MoE Framework}
\label{sec:method_moe}}

\begin{figure*}
  \includegraphics[width=1\textwidth]{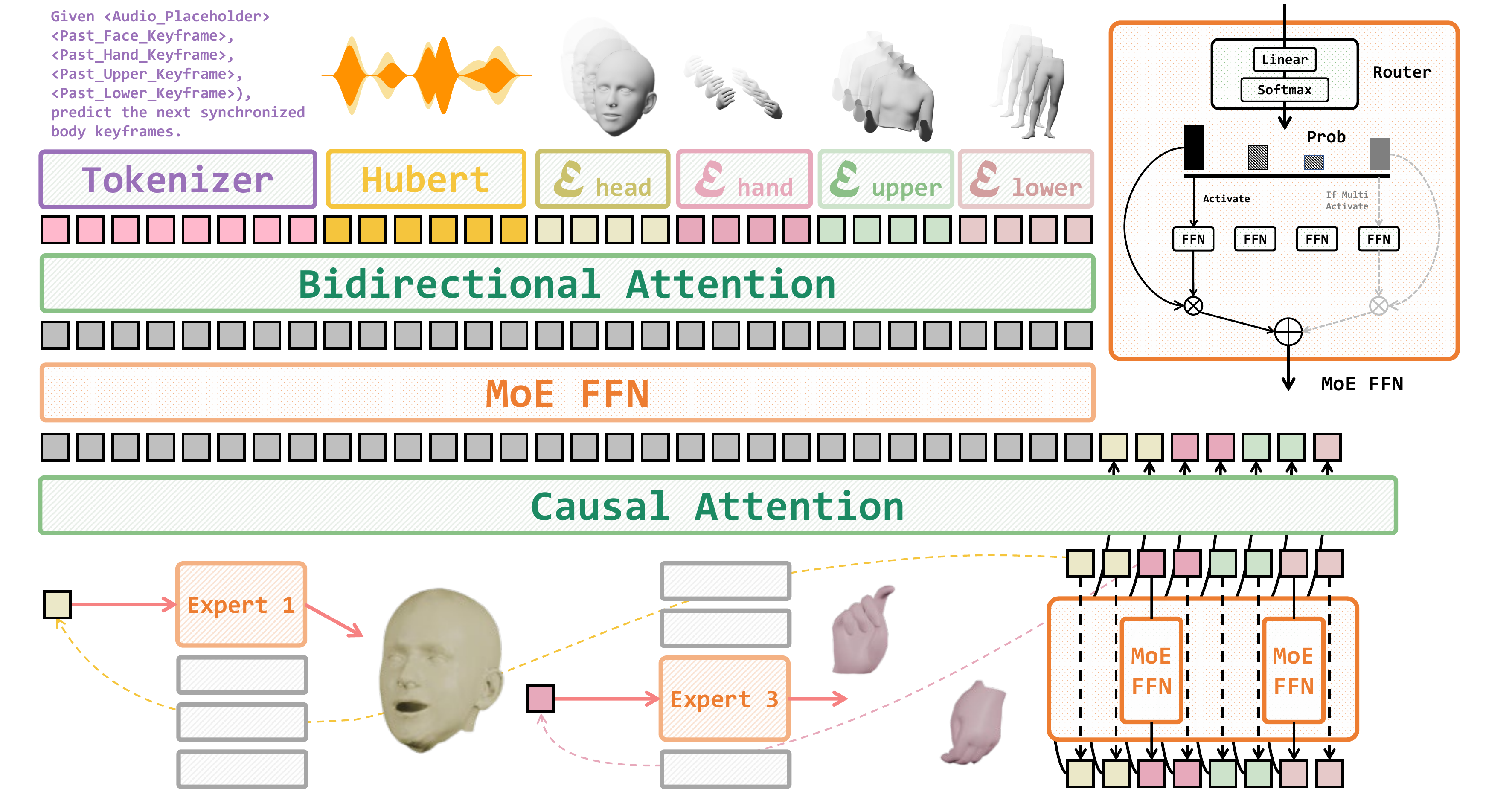}
  \caption{\textbf{MoE details}. The dense FFN layers within both the encoder and decoder are substituted with MoE FFNs. During inference, each token selectively activates a single expert from the four.}
  \label{fig:moe_detail}
\end{figure*}

As mentioned in Sec.~\mref{sec:related_work} and Sec.~\mref{sec:overview}, most of the preliminary work and we chose to decompose whole-body motion into face, hand, upper-body, and lower-body streams. A potential drawback of this design lies in the unique distributional patterns of motion features across different body parts. Therefore, utilizing shared parameters for inference may not only cause over-smoothing but also induce optimization conflicts due to gradient discrepancies. In practice, this results in diminished motion details and poor fitting of long-tail distributions.

UMo employs a Mixture of Experts (MoE) framework to address this issue (shown in Fig.~\mref{fig:moe_detail}). In our work, the MoE design offers two key advantages for real-time co-speech tasks by introducing spatial sparsity. First, it resolves the inconsistency in optimization objectives across different body parts. By routing tokens from specific parts to distinct experts, the model can better learn part-aware structured features while reducing optimization difficulty. Second, it enhances the model's latent representation capacity by increasing total parameters, yet avoiding the increase in computational overhead by controlling the number of activated routes. Instead of hard-partitioning the model into specific subnetworks, we use a soft specialization mechanism within a shared network. Here, routing dynamically allocates subnetworks based on contexts.

Our MoE framework starts from a pretrained dense T5 backbone and apply sparse upcycling to every feed-forward sub-module in both the encoder and decoder. For each transformer block we traverse all of its sub-layers and replace every FFN module with an MoE FFN, while leaving attention, layer normalization, and embeddings untouched. Each MoE module consists of a bias-free linear router $W_r\in\mathbb{R}^{d\times E}$ that maps token hidden states to expert logits and a bank of $E$ expert FFNs initialized as identical copies of the original dense FFN. Since all experts share the same initial weights, the MoE output is independent of the random router choice and functionally equivalent to the original dense model, allowing the router to learn specialization solely from the downstream objective.

Let $\mathbf{h}_t\!\in\!\mathbb{R}^{d}$ denote a hidden state arriving at an MoE FFN. The router yields logits $\mathbf{z}_t\!=\!W_r^{\top}\mathbf{h}_t$ and probabilities $\boldsymbol{\pi}_t\!=\!\mathrm{softmax}(\mathbf{z}_t)$. We pick the top-$k$ experts $\mathcal{S}_t\!\subseteq\!\{1,\dots,E\}$ and form
\begin{equation}
\begin{split}
\mathrm{MoE}(\mathbf{h}_t) &= \sum_{e\in\mathcal{S}_t} w_{t,e}\,\mathrm{FFN}_{e}(\mathbf{h}_t), \\
w_{t,e} &= \frac{\pi_{t,e}}{\max\!\left(\sum_{e'\in\mathcal{S}_t}\pi_{t,e'}\right)}.
\end{split}
\end{equation}
by default, we set $E\!=\!4$ and $k\!=\!1$. The in-set normalization degenerates to $w_{t,e}\!\equiv\!1$, so the layer reduces to deterministic top-$1$ routing and the router parameters receive gradients exclusively through the load-balancing auxiliary defined below. And we find this signal sufficient for stable specialization.

Since sparse routing can collapse onto a few overused experts. We therefore add a per-layer auxiliary $\ell_{\text{moe}}^{(\ell)}$ to the joint objective. Let $D^{(\ell)}_{t,e}\!\in\![0,1]$ denote the dispatched weight obtained by scattering the in-set-normalized top-$k$ probabilities of token $t$ back onto the full expert axis (so that $\sum_e D^{(\ell)}_{t,e}\!=\!1$ for every token), and let $N$ be the number of flattened tokens in the minibatch. We define
\begin{equation}
\label{lmoe}
\begin{split}
\ell_{\text{moe}}^{(\ell)} &= E\sum_{e=1}^{E} f_e^{(\ell)}\,\bar{\pi}_e^{(\ell)}, \\
f_e^{(\ell)} &= \frac{1}{N}\sum_{t=1}^{N} D^{(\ell)}_{t,e}, \quad
\bar{\pi}_e^{(\ell)} = \frac{1}{N}\sum_{t=1}^{N}\pi^{(\ell)}_{t,e}.
\end{split}
\end{equation}
Under our $k\!=\!1$ setting $D^{(\ell)}_{t,e}$ becomes a one-hot indicator and $f_e^{(\ell)}$ recovers the standard top-$1$ expert frequency, while the same expression remains valid for $k\!>\!1$. This term encourages balanced utilization while still permitting skewed routing when the data genuinely demands it.


During training, we additionally track the per-layer top-1 hit frequency and overall expert utilization to detect expert collapse early. At inference time, the auxiliary loss is disabled and only the sparse top-1 forward path is executed.

\begin{figure}[t]
  \includegraphics[width=0.5\textwidth]{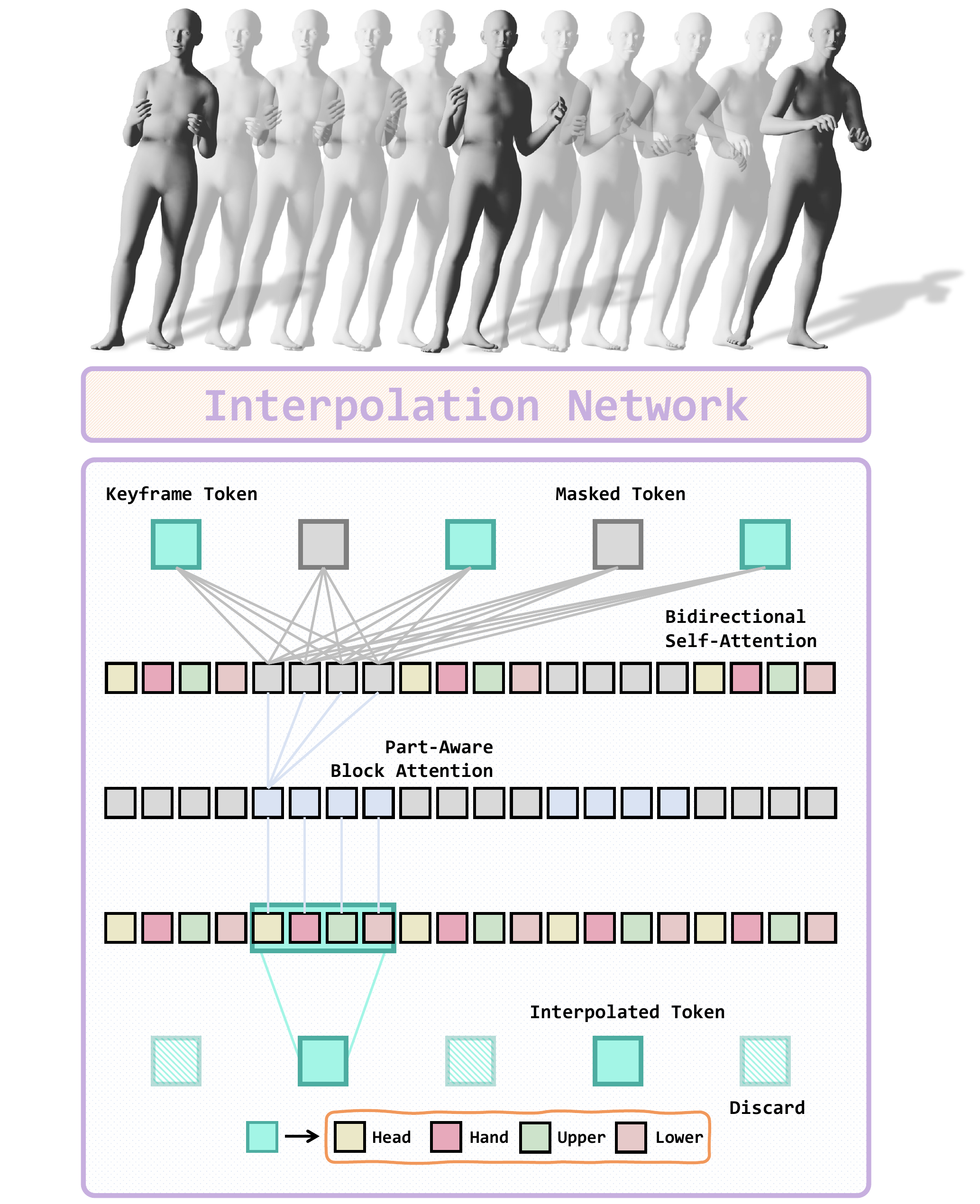}
  \caption{\textbf{Interpolation network}. Our interpolation network employs global bidirectional attention for temporal feature extraction and part-aware attention within frames to encode spatial features among parts. }
  \label{fig:Interpolation_network}
\end{figure}

\textbf{\subsection{Temporal Sparse Keyframe Framework}
\label{sec:method_interp}}

In real-time co-speech synthesis, achieving low-latency response is paramount for maintaining an immersive user experience. To effectively process complex linguistic semantics and generate high-fidelity motion sequences, we employ a T5-based architecture as our primary generative backbone. Although T5 is considered a relatively lightweight language model, its inference speed remains a significant bottleneck for strict real-time applications, particularly when the input encompasses high-dimensional multi-modal context.

To address this computational bottleneck, we introduce a temporal sparsity mechanism that decomposes the generation process into a two-stage pipeline. In the initial stage, the model synthesizes only a sparse set of keyframe poses to minimize temporal overhead. Subsequently, a more efficient and lightweight network is utilized to upsample these keyframes into a complete motion sequence in real time. This approach leverages the fact that dense per-frame generation is often redundant, since key semantic information in human motion resides in a few pivotal frames. Moreover, this design mimics professional animation pipelines, where artists focus on defining keyframes and rely on interpolation for smooth transitions.

Drawing inspiration from prior research that integrates keyframe abstractions into motion production \mcite{zheng2025autokeyframe}, we propose the keyframe framework to leverage this temporal sparsity for inference acceleration. By exploiting the inherent informational redundancy across consecutive frames, UMo achieves real-time throughput through an optimized interpolation network.

Before interpolation, sparse keyframe anchors are generated by the main UMo backbone. Given a window of $T$ frames and keyframe stride $s$ (default $s=6$ in our implementation), we define
\begin{equation}
\mathcal{K}=\{1,1+s,1+2s,\ldots\}\cap[1,T], \qquad K=|\mathcal{K}|,
\end{equation}
where $\mathcal{K}$ indexes keyframe positions and $K\ll T$.
Let $\mathbf{z}^{\text{kf}}=(z_{k_1},\ldots,z_{k_K})$ denote keyframe tokens over all motion parts at indices $\mathcal{K}$.
Conditioned on multimodal context $\mathbf{c}$ (audio/text/motion history), UMo performs sparse autoregressive decoding:
\begin{equation}
p_{\theta}(\mathbf{z}^{\text{kf}}\mid \mathbf{c})
=\prod_{i=1}^{K} p_{\theta}\!\left(z_{k_i}\mid z_{k_{<i}}, \mathbf{c}\right).
\end{equation}
Our framework shortens decoding length from $T$ to $K$, directly reducing latency while preserving long-range semantic control in the keyframe stream.

We optimize the interpolation network as a separate module, rather than jointly end-to-end with the UMo keyframe generator. This decoupling is motivated by the fact that keyframes predicted by the main backbone do not perfectly match the distribution of GT anchors. Directly enforcing joint optimization with GT supervision would couple interpolation targets with mismatched anchors, thereby violating the local smooth-transition assumption essential for interpolation. Consequently, the main backbone focuses on sparse keyframe generation, while the interpolation network learns anchor-constrained in-between reconstruction.

The interpolation network is trained using full token sequences where non-keyframe positions are masked. Specifically, indices in $\bar{\mathcal{K}}=[1,T]\setminus\mathcal{K}$ are replaced by mask tokens, while the keyframe indices $\mathcal{K}$ and the last valid frame serve as fixed anchors. The network is tasked with predicting only the masked positions, which are subsequently merged with the unchanged anchors to reconstruct the complete sequence.

Regarding the network architecture (shown in Fig.~\mref{fig:Interpolation_network}), the module first embeds each part stream using part-specific token embeddings. These are then aggregated by frame and processed via temporal self-attention. Following temporal encoding, a part-aware attention mechanism fuses information among the face, hand, upper, and lower body streams. Finally, part-specific prediction heads output logits for each stream. This ordering is designed to enhance both temporal stability and inter-part coordination.

For chunk-wise keyframe prediction,
\begin{equation}
p_\theta\!\left(\mathbf{z}^{\mathrm{kf}}_{t+P:t+L-1}\mid
\mathbf{z}^{\mathrm{kf}}_{t:t+P-1},\mathbf{a}_{t:t+L-1}\right),
\end{equation}
where $\mathbf{z}^{\mathrm{kf}}_{t:t+P-1}$ and $\mathbf{z}^{\mathrm{kf}}_{t+P:t+L-1}$ denote past and future keyframe-token chunks, respectively.
To reconstruct dense motion from sparse keyframes, the interpolation network builds a masked input sequence for each body part:
\begin{equation}
\tilde{z}_t=
\begin{cases}
z_t, & t\in\mathcal{K}\ \text{or}\ t=t_{\mathrm{last}},\\
z_{\mathrm{mask}}, & t\in\bar{\mathcal{K}},
\end{cases}
\end{equation}
where $t_{\mathrm{last}}$ is the last valid frame index in the current chunk.
Given $\tilde{\mathbf{z}}_{1:T}$ and frame-type tags, the interpolation network outputs logits $\mathbf{o}_t$ and predicts non-keyframe tokens by
\begin{equation}
\hat{z}_t=\arg\max_{v\in\mathcal{V}_{\mathrm{base}}}\mathbf{o}_{t,v},\qquad t\in\bar{\mathcal{K}}.
\end{equation}
The final complete motion sequence is constructed by merging the input keyframe anchors:
\begin{equation}
z_t^{\mathrm{rec}}=
\begin{cases}
z_t, & t\in\mathcal{K}\ \text{or}\ t=t_{\mathrm{last}},\\
\hat{z}_t, & t\in\bar{\mathcal{K}}.
\end{cases}
\end{equation}
In implementation, this procedure is applied in parallel to different body part and then returned as synchronized full-body token sequences.

\textbf{\subsection{Training Recipe and Audio Augmentation}\label{sec:method_training}}

\begin{figure}[t]
  \includegraphics[width=0.5\textwidth]{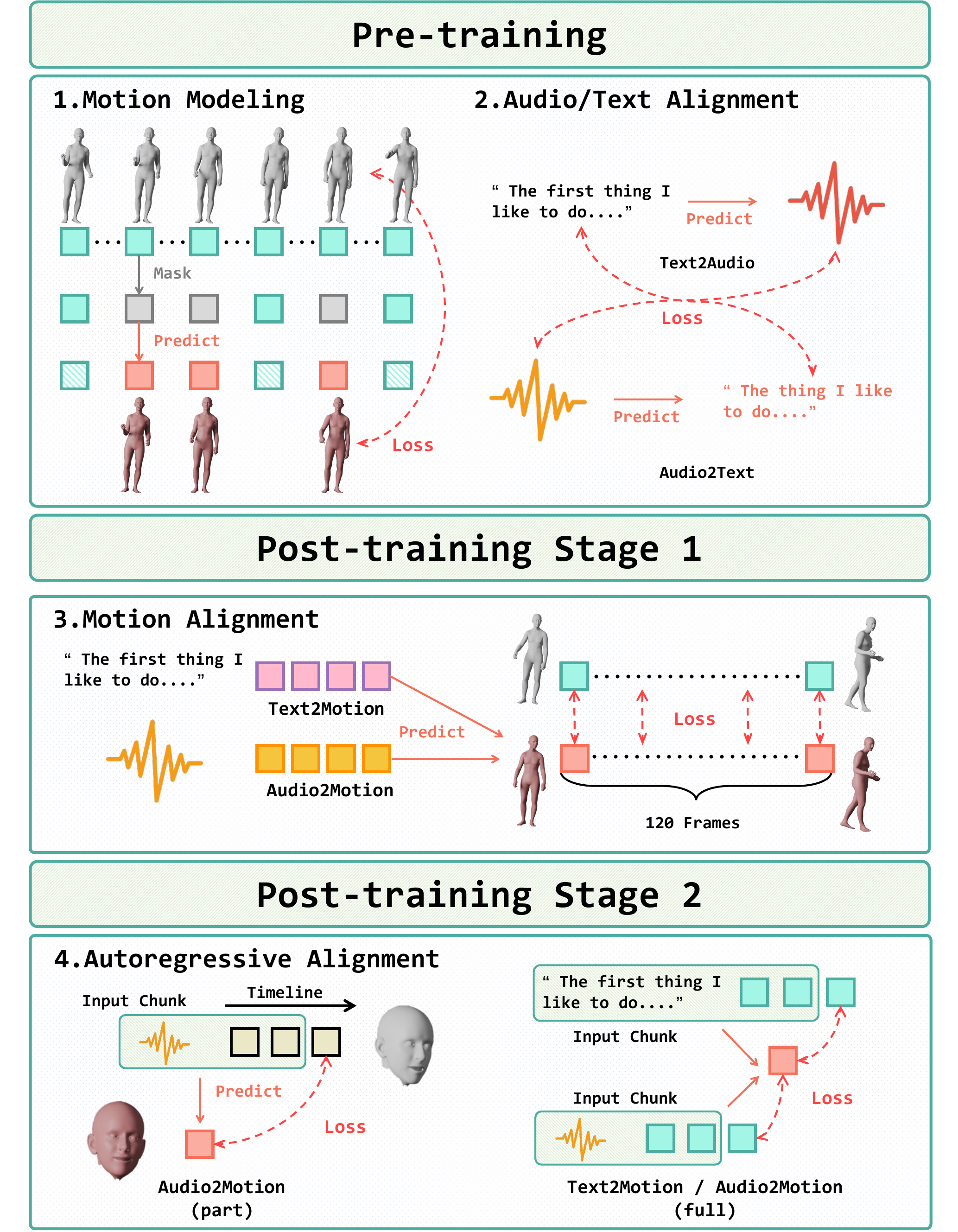}
  \caption{\textbf{Training recipe}. UMo integrates paired data of various formats and training objectives into a unified framework via a three-stage training recipe, enabling joint optimization for the co-speech task.}
  \label{fig:training_recipe}
\end{figure}

UMo uses a staged recipe to separate representation learning, modality alignment, and autoregressive ability (shown in Fig.~\mref{fig:training_recipe}).

To fully leverage the unified formulation, UMo minimizes task-conditional token prediction objectives across all modes under a shared compositional representation. This design not only simplifies model maintenance but also enables the seamless transfer of improvements across tasks. Consequently, the same representation, condition schema, and decoding constraints are preserved throughout pre-training, post-training, and online inference. For standard stages based on language models, we optimize token prediction over task-specific templates:
\begin{equation}
\mathcal{L}_{\text{LM}} = \mathbb{E}_{(x,y)\sim\mathcal{D}}[-\log p_{\theta}(y\mid x)].
\end{equation}

The unified objective represents as
\begin{equation}
\mathcal{L} =
\mathcal{L}_{\text{LM}}
+ \lambda_{\text{moe}}\,\mathcal{L}_{\text{moe}},
\label{eq:joint_objective}
\end{equation}
where $\lambda_{\text{moe}}$ is stage-dependent and explicitly set to $0$ during pre-training. The MoE loss $\mathcal{L}_{\text{moe}}$ is the average of $\ell_{\text{moe}}^{(\ell)}$ (described in Eq.~\mref{lmoe}) over all upcycled MoE layers.

\noindent\textbf{Pre-training.} 
In the pre-training stage, UMo draws samples from keyframe-based motion sequences (Sec.~\ref{sec:method_interp}) together with audio--text alignment pairs.
Each minibatch item is an instance $(x,y)$: encoder input $x$ packs instructions and multimodal placeholders, while decoder targets $y$ list discrete tokens for sparse keyframes and/or transcript/audio tokens depending on the template. All pre-training tasks (keyframe-based motion modeling, audio-to-text ($\mathrm{a2t}$) and text-to-audio ($\mathrm{t2a}$) alignment) share the same backbone and the same token-level cross-entropy objective, differing only in how their templates populate $x$ and $y$.

For the task-mixture sampling strategy, let $\mathcal{T}=\{t_1,\dots,t_N\}$ denote task families.
Each mini-batch is formed by first sampling a task $t\sim p(t)$ and then sampling a data instance from the corresponding task pool.
In practice, we use a temperature-controlled mixture
\begin{equation}
p(t_i)=\frac{n_i^{\tau}}{\sum_{j=1}^{N} n_j^{\tau}},
\label{eq:task_mixture}
\end{equation}
where $n_i$ is the available sample count for task $t_i$ and $\tau\in[0,1]$ controls balance between uniform mixing and data-proportional mixing.
This scheme avoids over-dominance of high-resource tasks while preserving exposure to dense generation data.

For each $t$ in pre-training data distribution $\mathcal{D}_{\mathrm{pre}}$,
\begin{equation}
\mathcal{L}_{\mathrm{pre}}
=
-\mathbb{E}_{t\sim p(t)}\,\mathbb{E}_{(x,y)\sim\mathcal{D}_{\mathrm{pre}}}
\sum_{j\in\mathcal{S}_{\mathrm{sup}}(y)}
\log p_{\theta}\!\left(y_j \mid y_{<j}, x\right),
\end{equation}
where $\mathcal{S}_{\mathrm{sup}}(y)\subseteq\{1,\ldots,|y|\}$ denote positions with valid supervision. The relative contribution of each task is controlled implicitly by its sampling ratio in $\mathcal{D}_{\mathrm{pre}}$ rather than by an explicit loss-weight hyperparameter. 
This formulation corresponds to the unified objective in Eq.~\mref{eq:joint_objective} with $\lambda_{\text{moe}}{=}0$, so that the MoE auxiliary loss remains inactive until post-training.

\noindent\textbf{Post-training Stage~1.} 
Stage~1 initializes from the pre-trained checkpoint and performs instruction tuning to strengthen motion-audio alignment under dense temporal supervision, where every frame within the window contributes to the loss. In addition to the primary audio-to-motion ($\mathrm{a2m}$) templates, we mix in text-to-motion ($\mathrm{t2m}$) templates as an auxiliary task to regularize $\mathrm{a2m}$ alignment with linguistic priors from pre-training.

Let $\mathcal{D}_{\mathrm{S1}}$ denote Stage~1 templates and $\mathcal{F}_{\mathrm{S1}}(x,y)\subseteq\mathcal{S}_{\mathrm{sup}}(y)$ the supervised decoder indices.
The Stage~1 loss augments token prediction with sparse MoE regularization:

\begin{equation}
\begin{split}
\mathcal{L}_{\mathrm{S1}}
= -\mathbb{E}_{t\sim p(t)}\,\mathbb{E}_{(x,y)\sim\mathcal{D}_{\mathrm{S1}}}
\sum_{j\in\mathcal{F}_{\mathrm{S1}}(x,y)}
\log p_{\theta}\!\left(y_j \mid y_{<j}, x\right) \\
+ \lambda_{\mathrm{moe}}\mathcal{L}_{\mathrm{moe}}.\qquad\qquad
\end{split}
\end{equation}

At this stage, the motion window per step contains $120$ keyframes.

\noindent\textbf{Post-training Stage~2.} 
Stage~2 inherits the $\mathrm{a2m}$/$\mathrm{t2m}$ task mixture and motion window length from Stage~1, and further mixes full-body with part-level keyframe objectives for chunk-wise autoregression.

Let $\mathcal{F}_{\mathrm{S2}}(x,y)$ mark eligible decoder positions (excluding prefixes, padding, and task-specific holdouts).
We optimize
\begin{equation}
\begin{split}
\mathcal{L}_{\mathrm{S2}}
= -\mathbb{E}_{t\sim p(t)}\,\mathbb{E}_{(x,y)\sim\mathcal{D}_{\mathrm{S2}}}
\sum_{j\in\mathcal{F}_{\mathrm{S2}}(x,y)}
\log p_{\theta}\!\left(y_j \mid y_{<j}, x\right) \\
+ \lambda_{\mathrm{moe}}\mathcal{L}_{\mathrm{moe}}.\qquad\qquad
\end{split}
\end{equation}
Sparse keyframe streams in Stage~2 follow the autoregressive framework introduced in Sec.~\mref{sec:autoregressive_design} and Sec.~\mref{sec:method_interp}, where the backbone predicts only keyframe tokens at stride $s$. The complete motion sequence reconstruction is delegated to the interpolation network.

\noindent\textbf{Interpolation Training.} 
For interpolation network, we optimize $\mathcal{L}_{\text{interp}}$.
Given sparse keyframes, we mask non-keyframe positions and predict only masked tokens. Keyframes remain fixed as hard constraints.

Let $m_t\in\{0,1\}$ indicate whether frame $t$ is masked, and let $\mathcal{M}=\{t : m_t{=}1\}=\bar{\mathcal{K}}$ collect the masked positions. Let $\mathbf{c}$ collect the auxiliary conditions used by the interpolation network. We deliberately remove raw audio from $\mathbf{c}$ so that the interpolation network focuses on in-betweening smoothness between fixed keyframes. For each step, the cross-entropy loss is
\begin{equation}
\mathcal{L}_{\text{ce}} =
- \sum_{t\in\mathcal{M}} \log p_{\theta}\!\left(z_t \,\middle|\, z_{\mathcal{K}}, \mathbf{c}\right),
\end{equation}
where $z_{\mathcal{K}}$ are the anchor keyframe tokens. Both keyframe anchors and padding positions are excluded from CE aggregation.

To discourage abrupt jumps between adjacent predictions and to shape the per-part token embedding table into a locally continuous geometry, we add a light-weight smoothness regularizer defined directly on argmax-decoded embedding trajectories.
Concretely, for each part stream $\pi\in\{\text{face},\text{hand},\text{upper},\text{lower}\}$ we take the argmax over the full output sequence, look the predicted ids up in the part-specific token embedding table, and obtain a per-frame embedding sequence $\{\mathbf{e}^{(\pi)}_t\}$.
The argmax operation is detached from the computation graph, so gradients flow only through the token-embedding lookup; the regularizer therefore acts as a continuity prior on the embedding table itself rather than as a direct penalty on predicted logits.
We then apply masked first- and second-order finite differences that skip any pair (resp.\ triple) touching a padded position, and average over valid positions and parts:
\begin{equation}
\mathcal{L}_{\text{vel}} =
\frac{1}{|\Pi|}\sum_{\pi\in\Pi}
\frac{1}{N^{(\pi)}_{\text{vel}}}
\sum_{(t-1,t)\in\mathcal{P}^{(\pi)}_{\text{vel}}}
\bigl\lVert \mathbf{e}^{(\pi)}_{t}-\mathbf{e}^{(\pi)}_{t-1}\bigr\rVert_2^2,
\end{equation}
\begin{equation}
\mathcal{L}_{\text{acc}} =
\frac{1}{|\Pi|}\sum_{\pi\in\Pi}
\frac{1}{N^{(\pi)}_{\text{acc}}}
\sum_{(t-1,t,t+1)\in\mathcal{P}^{(\pi)}_{\text{acc}}}
\bigl\lVert \mathbf{e}^{(\pi)}_{t+1}-2\mathbf{e}^{(\pi)}_{t}+\mathbf{e}^{(\pi)}_{t-1}\bigr\rVert_2^2,
\end{equation}
where $\mathcal{P}^{(\pi)}_{\text{vel}}$ and $\mathcal{P}^{(\pi)}_{\text{acc}}$ collect index pairs and triples whose neighbors are all non-padded, and $N^{(\pi)}_{\text{vel}}$, $N^{(\pi)}_{\text{acc}}$ are their cardinalities scaled by the embedding dimension.
The final interpolation objective combines CE and smoothness terms:
\begin{equation}
\mathcal{L}_{\text{interp}} =
\mathcal{L}_{\text{ce}}
+ \lambda_v \mathcal{L}_{\text{vel}}
+ \lambda_a \mathcal{L}_{\text{acc}}.
\end{equation}

\noindent\textbf{Inference.} 
During inference, we roll out keyframe-based autoregression in a sliding-window manner.
Each decoding step instantiates a task template $\tau$ (instruction text with modality placeholders), the aligned audio segment $a$ for the current temporal window, and the most recent $P{=}10$ committed keyframe tokens per body stream as motion history $\mathbf{z}^{\mathrm{kf}}_{\mathrm{hist}}$.
The backbone predicts the next $N{=}5$ keyframe steps for face, hands, upper body, and lower body jointly.
After which the newly generated keyframes are appended, the buffer is shifted forward, and the procedure repeats for online continuation.

To reconstruct complete motion from sparse keyframes, the committed keyframes are placed at their anchor indices $\mathcal{K}$ and all non-keyframe positions in the current window are filled with the mask token.
The interpolation network then predicts the masked tokens in a single forward pass while treating keyframe anchors as hard constraints, and the predictions are merged with the anchors via anchor-preserving substitution.
Finally, generated part tokens are decoded into continuous motion parameters, merged into a full-body sequence, and converted to SMPL-X outputs.

\begin{table*}[t]
\centering
\caption{Comparisons on BEATv2 benchmark (speaker-2). We follow the benchmark convention \cite{liu2024emage, chen2024language, xu2024mambatalk} and report FGD$\times 10^{-1}$, BC$\times 10^{-1}$, and Diversity. The FPS and TTFF is computed by one NVIDIA A100. We color the \colorbox{tab1!70}{best}, \colorbox{tab2!70}{second-best}, and \colorbox{tab3!70}{third-best} results. * data is calculated proportionally from other papers.}
\label{tab:exp_main}
\scalebox{0.87}{
\begin{tabular}{lcccccccc}
\midrule[0.8mm]
Method & Unified & Autoregressive & FGD$\times 10^{-1}$ ($\downarrow$) & BC$\times 10^{-1}$ ($\uparrow$) & Diversity ($\uparrow$) & FPS ($\uparrow$) & TTFF (ms) ($\downarrow$) \\
\midrule[0.5mm]
\midrule[0.5mm]
\multicolumn{7}{l}{\textit{Specific Methods}} \\
\midrule[0.3mm]
Habibie \textit{et al.} \mcite{habibie2021learning} & - & - & 9.040 & 7.716 & 8.043 & - & - \\
DisCo \mcite{liu2022disco} & - & - & 9.417 & 6.439 & 9.912 & - & - \\
CaMN \mcite{liu2022beat} & - & - & 6.644 & 6.769 & 10.86 & - & - \\
TalkShow \mcite{yi2023generating} & - & - & 6.209 & 6.947 & 13.47 & - & - \\
EMAGE \mcite{liu2024emage} & - & - &  5.512 & 7.724 & 13.06 & - & - \\
SynTalker \mcite{chen2024enabling} & \checkmark & - & 6.413 & \colorbox{tab1!70}{7.971} & 12.72 & - & - \\
SynTalker (Audio only) & - & - & \colorbox{tab2!70}{4.687} & 7.363 & 12.43 & $6^*$ & - \\
MambaTalk \mcite{xu2024mambatalk} & - & - & 5.366 & 7.812 & 13.05 & 1555 & 38 \\
RAG-Gesture \mcite{mughal2025retrieving} & - & - & 8.790 & 7.300 & 12.62 & - & - \\
GestureLSM \mcite{liu2025gesturelsm} & - & - & \colorbox{tab1!70}{4.247} & 7.290 & 13.76 & 411 & 275 \\
\midrule[0.5mm]
\multicolumn{7}{l}{\textit{LLM-based Methods}} \\
\midrule[0.3mm]
LOM \mcite{chen2024language} & \checkmark & \checkmark &  5.301 & 7.780 & 15.17 & 19 & 6269 \\
MIBURI \mcite{mughal2026miburi} & - & \checkmark & 7.530 & 7.900 & \colorbox{tab3!70}{15.85} & $57^*$ & $35^*$ \\
\textbf{UMo (Ours)} & \checkmark & \checkmark & \colorbox{tab3!70}{5.107} & \colorbox{tab2!70}{7.955} & 14.75 & 44 & 826 \\
- w. Audio Aug. & \checkmark & \checkmark & 5.368 & 7.787 & \colorbox{tab1!70}{16.77} & - & - \\
- w. Audio Aug. (N=2) & \checkmark & \checkmark & 5.464 & \colorbox{tab3!70}{7.919} & \colorbox{tab2!70}{16.21} & 27 & 540 \\
\midrule[0.8mm]
\end{tabular}}
\end{table*}

\noindent\textbf{Audio Augmentation.} Current co-speech synthesis frameworks rely on datasets with a strict one-to-one alignment between audio and motion. Although this alignment is ideal, acquiring such synchronized data is costly, resulting in significantly smaller training datasets compared to standard motion datasets. This data scarcity creates two main challenges, including a lack of robustness to diverse inputs and limited motion-audio mapping diversity.

Unlike linguistic text, audio signals contain unique timbral signatures that vary across speakers. Although state-of-the-art codecs like HuBERT \mcite{hsu2021hubert} attempt to disentangle timbre from semantic content, the resulting latent representations often remain partially entangled. Consequently, when input audio deviates from the training distribution, the network struggles to generalize to these unseen variations. Furthermore, it is crucial to distinguish between motion primitives conveying universal semantics and those reflecting personal traits. In scenarios where data is limited, capturing broader motion primitives should take precedence over learning personal features.

To overcome the limitations of strict one-to-one alignments under limited data, we adopt an audio augmentation strategy to facilitate the learning of universal motion patterns. Specifically, we use a Text-to-Speech model to generate five distinct audio variants for each motion sequence, each with a unique vocal timbre. The computational cost of this synthesis is negligible compared to the significant overhead of capturing new co-speech motion data. By establishing a many-to-one mapping from diverse audio inputs to a single motion output, we encourage the network to distill semantic commonalities while enhancing its robustness to out-of-distribution acoustic features.

\textbf{\section{Experiments}}
\label{sec:exp}

\begin{table}[t]
\centering
\caption{User Study. Participants are requested to rank the outputs of the five methods. We report the average rank in the table, where a lower value indicates better performance.}
\label{tab:exp_user}
\scalebox{0.85}{
\begin{tabular}{cccc}
\midrule[0.8mm]
Method & Smoothness ($\downarrow$) & Diversity ($\downarrow$) & Preference ($\downarrow$) \\
\midrule[0.5mm]
LOM & \colorbox{tab3!70}{3.17} & \colorbox{tab3!70}{2.96} & 3.35 \\
MambaTalk & 3.56 & 3.45 & 3.36 \\
GestureLSM & \colorbox{tab2!70}{2.72} & 4.00 & \colorbox{tab3!70}{3.26} \\
\midrule[0.3mm]
UMo (Ours) & 3.54 & \colorbox{tab2!70}{2.34} & \colorbox{tab2!70}{3.23} \\
- w. Audio Aug. & \colorbox{tab1!70}{1.99} & \colorbox{tab1!70}{2.22} & \colorbox{tab1!70}{1.78} \\
\midrule[0.8mm]
\end{tabular}}
\end{table}

\begin{figure*}[t]
  \includegraphics[width=0.97\textwidth]{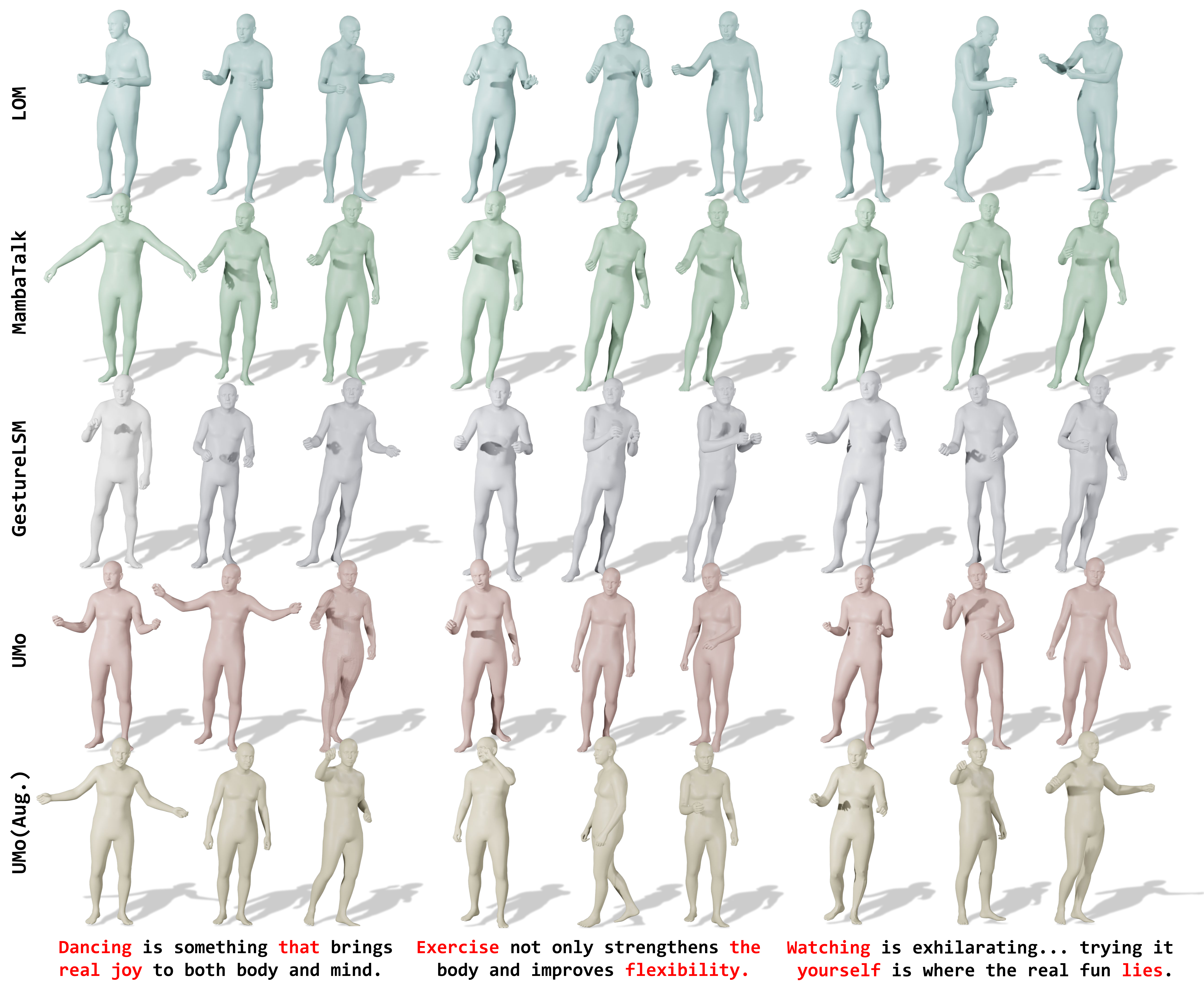}
  \caption{\textbf{Qualitative Comparisons}. We visualize the motion sequence of LOM \mcite{chen2024language}, MambaTalk \mcite{xu2024mambatalk}, GestureLSM \mcite{liu2025gesturelsm} and UMo with the same audio input. Our results are more vivid and reasonable compared to other methods. Aug. here means audio augmentation. }
  \label{fig:comparisons}
\end{figure*}

\textbf{\subsection{Experimental Setup}}
\label{sec:exp_setup}

\noindent\textbf{Datasets.}
We evaluate co-speech gesture generation on BEATv2 \mcite{liu2022beat, liu2024emage}. BEATv2 includes 60 hours co-speech data collected from 25
speakers (12 female, 13 male). Following previous work \mcite{liu2024emage, chen2024language}, we train and test on speaker-2 for the primary benchmark results.

\noindent\textbf{Metrics.}
We report Fr\'echet Gesture Distance (FGD), Beat Constancy (BC), and $\ell_1$ diversity, following standard practice on this benchmark. To evaluate inference speed, we measure end-to-end wall-clock latency on one NVIDIA A100 GPU, reporting the results in Frames Per Second (FPS) and Time To First Frame (TTFF).

\noindent\textbf{Implementation Details.} We report the default settings used in the main training and testing. We optimize the model with AdamW using a learning rate of $1\times 10^{-4}$, $\beta_1=0.9$, $\beta_2=0.99$, and a weight decay of $0.05$. All experiments are conducted in bfloat16 precision with a batch size of $24$. We train the full model on $8\times$ NVIDIA A100-80GB GPUs, and each configuration is trained until convergence, which takes approximately $150$ epochs with a wall-clock time of around one hour per epoch.

For motion tokenization, we directly adopt the pre-trained VQ-VAE weights from LOM~\mcite{chen2024language} and keep them frozen throughout all training stages. For T5-MoE, we set the number of experts to $4$ with top-$k$ routing ($k=1$), and an MoE auxiliary loss weight of $\lambda_{\text{moe}}=0.01$. For the interpolation network, we use a keyframe stride of $s=6$, label smoothing of $0.1$, a velocity regularization weight $\lambda_v=10^{-3}$, and an acceleration regularization weight $\lambda_a=10^{-4}$. To improve robustness to variable context lengths, we apply short-context augmentation during sample construction, which stabilizes generation when only limited history is available. For audio enhancement, we adopt minimax-speech-2.6~\mcite{minimax_speech_2025}. During testing, we pre-fill the $10$-frame prefix history of the first chunk with the GT keyframe.

\begin{figure*}[t]
  \includegraphics[width=1\textwidth]{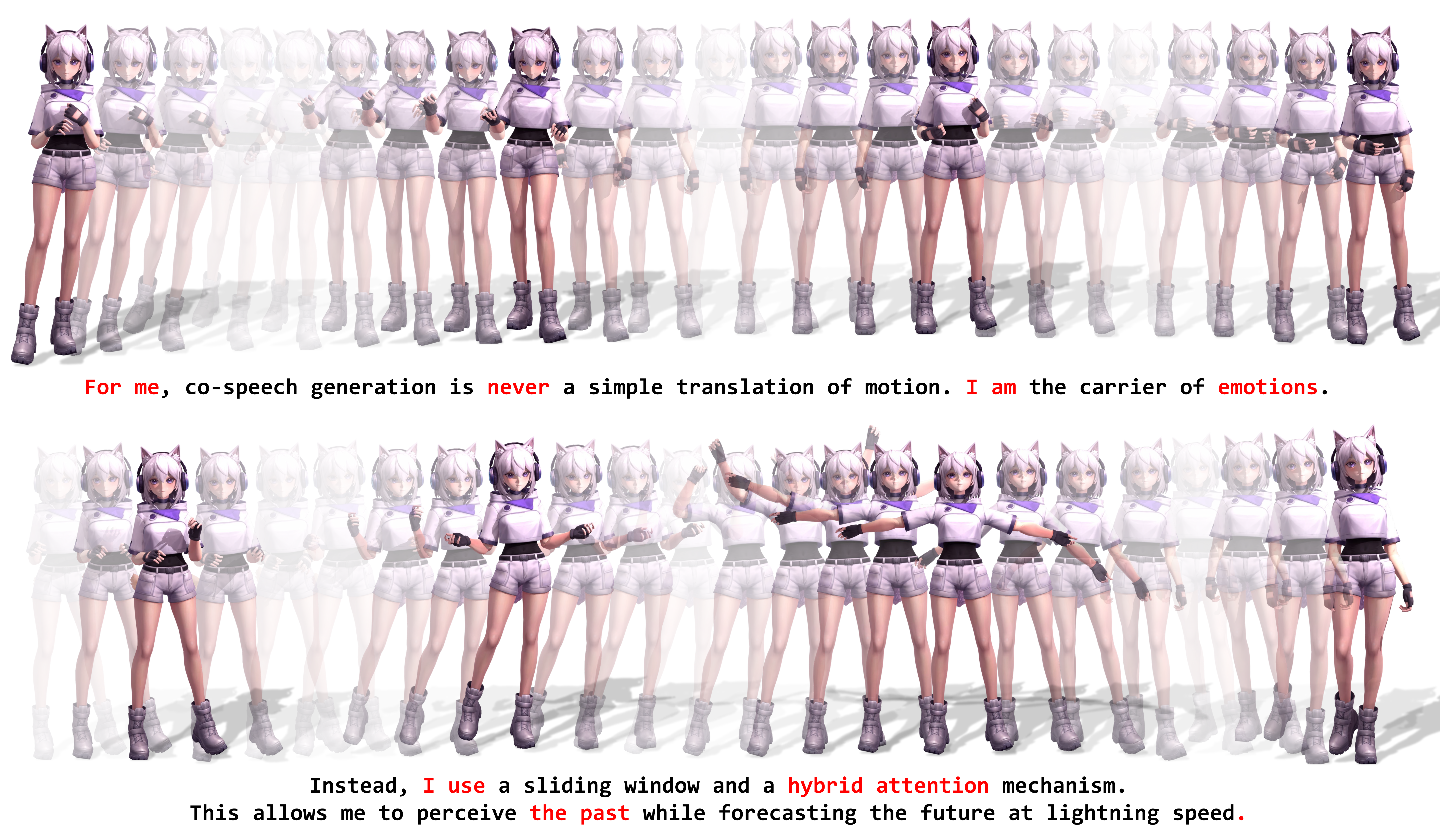}
  \caption{\textbf{Visualization Results of UMo}. The model is trained on BEATV2 females and the SMPLX motion sequences are retargeted to the avatar. We mainly showcase the rhythm and expressiveness of the movements, but the expressions are not synchronized. }
  \label{fig:ours_more}
\end{figure*}

\textbf{\subsection{Quantitative Experiments}\label{sec:exp_quant}}

Table~\mref{tab:exp_main} summarizes co-speech gesture generation on BEATv2 (speaker-2) together with wall-clock statistics measured on one NVIDIA A100 GPU.

\begin{table*}[t]
\centering
\caption{Ablation study on UMo components, including autoregressive(Sec.~\mref{sec:autoregressive_design}), MoE (Sec.~\mref{sec:method_moe}), keyframe (Sec.~\mref{sec:method_interp}), and audio augmentation (Sec.~\mref{sec:method_training}). We color the \colorbox{tab1!70}{best}, \colorbox{tab2!70}{second-best}, and \colorbox{tab3!70}{third-best} results. }
\label{tab:ablation_frame}
\scalebox{0.93}{
\begin{tabular}{lcccccccc}
\midrule[0.8mm]
Method & \hspace{1em}AR\hspace{1em} & MoE & Keyframe & Interp. Network & Audio Aug. & FGD$\times 10^{-1}$ ($\downarrow$) & BC$\times 10^{-1}$ ($\uparrow$) & Diversity ($\uparrow$) \\
\midrule[0.5mm]
\midrule[0.5mm]
\multicolumn{4}{l}{\textit{Baseline}} \\
\midrule[0.5mm]
- & - & - & - & - & - & 5.81 & 7.97 & 15.73 \\
\midrule[0.5mm]
\multicolumn{4}{l}{\textit{Variants}} \\
\midrule[0.5mm]
Var. 1 & \checkmark & - & - & - & - & 6.810 & \colorbox{tab1!70}{8.117} & 15.60 \\
Var. 2 & - & \checkmark & - & - & - & 5.767 & \colorbox{tab2!70}{7.989} & \colorbox{tab3!70}{16.18} \\
Var. 3 & - & - & \checkmark & - & - & 5.607 & 7.866 & 14.63 \\
Var. 4 & - & - & \checkmark & \checkmark & - & \colorbox{tab3!70}{5.196} & 7.790 & 14.55 \\
Var. 5 & \checkmark & \checkmark & - & - & - & 6.434 & 7.608 & 14.79 \\
Var. 6 & - & \checkmark & \checkmark & - & - & 5.578 & 7.936 & 15.02 \\
Var. 7 & - & \checkmark & \checkmark & \checkmark & - & \colorbox{tab2!70}{5.137} & 7.834 & 14.88 \\
Var. 8 & \checkmark & - & \checkmark & \checkmark & - & 5.331 & 7.899 & 14.72 \\
Var. 9 & \checkmark & \checkmark & \checkmark & \checkmark & - & \colorbox{tab1!70}{5.107} & \colorbox{tab3!70}{7.955} & 14.75 \\
Var. 10 (N=5) & \checkmark & \checkmark & \checkmark & \checkmark & \checkmark & 5.368 & 7.787 & \colorbox{tab1!70}{16.77} \\
Var. 11 (N=2) & \checkmark & \checkmark & \checkmark & \checkmark & \checkmark & 5.464 & 7.919 & \colorbox{tab2!70}{16.21} \\
\midrule[0.8mm]
\end{tabular}}
\end{table*}

Compared with LOM \mcite{chen2024language}, UMo improves both fidelity and synchronization while substantially reducing interaction latency. The MoE framework in UMo equips the model with superior representation capabilities. Meanwhile, leveraging historical context via keyframes ensures the coherence of the overall motion.

In comparison to MIBURI \mcite{mughal2026miburi}, UMo exhibits a higher TTFF. This is primarily because, although MIBURI employs a duplex audio model to handle user input and provide audio responses, its motion generation only relies on a lightweight network. In contrast, UMo routes co-speech motion through a unified LLM backbone, meaning motion tokens are produced by the same LM inference stack as multimodal conditioning. While this design incurs a higher TTFF than MIBURI, it yields exceptional motion expressiveness, achieving the best FGD among LLM-based methods. Furthermore, reducing the autoregressive window during inference from $N=5$ to $N=2$ decreases the TTFF. However, this reduction is not linear, and the latency overhead independent of the window size ultimately leads to a drop in the overall FPS.

LLM-based generators achieve the highest diversity values in the table compared with task-specific methods, whereas UMo ranks slightly lower. We attribute this gap to UMo's keyframe design, where the primary motion sequence is generated by an in-betweening network. Consequently, with a fixed total size, only a minority of key anchor points are outputs from the LLM backbone, which inevitably results in slightly lower overall motion diversity. While our proposed audio augmentation method significantly enhances the diversity and substantially improves motion expressiveness. The observed drop in FGD is expected, as this strategy is not specifically optimized for the audio distribution of the test data.

Beyond these nearest competitors, Table~\mref{tab:exp_main} situates UMo relative to a broader family of co-speech methods.
Compared with representative methods such as EMAGE \mcite{liu2024emage}, and MambaTalk \mcite{xu2024mambatalk}, UMo simultaneously lowers FGD and raises BC, suggesting that the unified token LM backbone plus sparse keyframe decoding improves both perceptual realism and beat alignment relative to older pipelines. While compared with SynTalker \mcite{chen2024enabling}, in the mode of training with unified tasks simultaneously, UMo performs much better than Syntalker (FGD 5.107 verse 6.413). However, when focusing on audio2motion, Syntalker has better performance in FGD (4.687), but its inference speed falls far short of real-time requirements. While UMo outperforms GestureLSM in terms of BC and Diversity, it still lags behind GestureLSM in FGD and speed metrics. This phenomenon can be combined with The results of Syntalker, which also indirectly indicate that under the same data conditions, a unified task optimization method may have a certain negative effect on the indicators of a single task.

Additionally, Table~\mref{tab:exp_main} situates UMo relative to a broader family of co-speech methods. Compared with representative methods such as EMAGE \mcite{liu2024emage} and MambaTalk \mcite{xu2024mambatalk}, UMo simultaneously lowers FGD and raises BC. This suggests that the Moe framework, combined with sparse keyframe decoding, improves both perceptual realism and beat alignment relative to older pipelines. In the unified multi-task training setting, UMo significantly outperforms SynTalker \mcite{chen2024enabling} (FGD 5.107 vs. 6.413). However, when training solely on audio-to-motion, SynTalker achieves a better FGD (4.687), yet its inference speed falls far short of real-time requirements.
While UMo outperforms GestureLSM in terms of BC and Diversity, it still lags behind GestureLSM in FGD and speed metrics. This trend, consistent with the SynTalker results, indirectly indicates that under identical data conditions, a unified task optimization approach may incur a slight trade-off in the performance metrics of individual tasks.

Taken together, the table positions UMo as closing much of the fidelity gap to strong non-LLM models while preserving the deployment advantages of sparse decoding.

\begin{table}[t]
\centering
\caption{Ablation study on training stages.}
\label{tab:exp_ablation_recipe}
\scalebox{0.75}{
\begin{tabular}{lcccccc}
\midrule[0.8mm]
Method & FGD$\times 10^{-1}$ ($\downarrow$) & BC$\times 10^{-1}$ ($\uparrow$) & Diversity ($\uparrow$) \\
\midrule[0.5mm]
- w/o. Pre-train & 5.920 & 6.572 & 15.03\\
- w/o. Post-train Stage 1 & 5.422 & 7.416 & \colorbox{tab1!70}{15.41} \\
- w/o. Post-train Stage 2 & 5.137 & 7.834 & 14.88 \\
\textbf{UMo} & \colorbox{tab1!70}{5.107} & \colorbox{tab1!70}{7.955} & 14.75 \\
\midrule[0.8mm]
\end{tabular}}
\end{table}

\textbf{\subsection{Qualitative Evaluation}\label{sec:exp_quali}}
The score on BEATv2 mainly reflects the alignment with the original data. However, the score fails to reflect the expressiveness of the motion, and it also doesn't accurately capture the audio-motion synchronization. We therefore conduct the qualitative evaluation and user study.

We show the visualization results in Fig.~\mref{fig:comparisons} with out-of-domain audios. We compare UMo against recent strong baselines with publicly available implementations or official demos (all models are trained on Speaker 2 only), including LOM \mcite{chen2024language}, MambaTalk \mcite{xu2024mambatalk}, and GestureLSM \mcite{liu2025gesturelsm}, rendered under identical skeleton for fair visual comparison. Across these examples, MambaTalk and GestureLSM produce relatively smooth motions, but they suffer from limited diversity and fail to fully adapt to the variations in audio input. LOM exhibits better diversity, yet its alignment with the audio remains suboptimal. In contrast, UMo demonstrates significantly superior motion diversity and expressiveness compared to other methods. It also tends to preserve sharper phrase-level gestures aligned with emphasis and maintains coherent hand--face coordination, especially when the audio contains fast-paced syllables. Cases where beat trackers disagree with subjective rhythm are retained to illustrate remaining ambiguity. We further present the extra results of UMo in Fig.~\mref{fig:ours_more}. The SMPLX sequence is redirected onto the character model to provide a more realistic sense of expressiveness.

The user study is also conducted with anonymized side-by-side videos. We generated 10 video clips, each lasting 10 to 20 seconds, for 5 different methods using the standard SMPLX mesh. Each method is trained on BEATv2 (Speaker2) and accpet the same out-of-domain audio as input. We then concatenated the motion sequences from all five methods into a single video for each sample, with the order of the methods randomly shuffled. Subsequently, we invited 13 participants to rank the motion sequences based on their smoothness, diversity, and human preference after viewing each pair. Aggregated statistics are reported in Table~\mref{tab:exp_user}. The results show that UMo aligns better with overall user preferences. Without audio augmentation, UMo exhibits slightly inferior motion continuity, yet its motion diversity remains outstanding. Despite its high smoothness, GestureLSM lacks motion diversity, which makes it less favored by users. With audio augmentation, UMo significantly outperforms other methods.

\textbf{\subsection{Ablation Studies}\label{sec:exp_ablation}}

In this section, we conduct ablation studies on each module of UMo to verify its effectiveness. Specifically, we analyze the function of UMo components, training recipe, keyframe stride, interpolation, and runtime.

\noindent\textbf{Components Analysis.} We report the complete component ablation results in Table~\mref{tab:ablation_frame}. The baseline is obtained by removing all components and training/testing solely with BEATv2 (Speaker2). We thoroughly test the combination of our AR (Sec.~\mref{sec:autoregressive_design}), MoE (Sec.~\mref{sec:method_moe}), and keyframe (Sec.~\mref{sec:method_interp}) frameworks. In non-AR scenarios, the prediction window size is 120 frames. Additionally, audio augmentation is applied exclusively during the post-training Stage 2 of our final version.

Experiments show that using AR alone significantly degrades generation quality despite the gain in response speed (Var.1). This indicates that simply shortening the decoding horizon is detrimental. However the negative impact of AR on quality can be mitigated by MoE and the keyframe strategy. MoE directly enhances generation quality within a single inference pass through multi-expert decoding (Var.5). The positive effect of the keyframe strategy on AR (Var.8) likely stems from an implicit window expansion. Given the same number of input frames keyframes provide a longer temporal receptive field by skipping adjacent frames with low information density thereby improving the overall quality of motion generation. 

\begin{table}[t]
\centering
\caption{Ablation study on keyframe stride for motion quality (N=5).}
\label{tab:exp_ablation_keyframe}
\scalebox{0.82}{
\begin{tabular}{cccc}
\midrule[0.8mm]
Keyframe Interval & FGD$\times 10^{-1}$ ($\downarrow$) & BC$\times 10^{-1}$ ($\uparrow$) & Diversity ($\uparrow$) \\
\midrule[0.5mm]
\midrule[0.5mm]
\multicolumn{4}{l}{\textit{Baseline}} \\
\midrule[0.5mm]
- & 5.81 & 7.97 & \colorbox{tab1!70}{15.73}\\
\midrule[0.5mm]
\multicolumn{4}{l}{\textit{Linear Interpolation}} \\
\midrule[0.5mm]
s = 4 & 5.84 & 8.15 & 14.53\\
s = 6 & 5.90 & 8.01 & 14.37\\
s = 8 & 6.04 & 8.11 & 14.63 \\
\midrule[0.5mm]
\multicolumn{4}{l}{\textit{Network Interpolation}} \\
\midrule[0.5mm]
s = 4 & 5.63 & \colorbox{tab1!70}{8.11} & 14.29\\
s = 6 & \colorbox{tab1!70}{5.107} & 7.955 & 14.75\\
s = 8 & 5.12 & 7.85 & 14.31\\
\midrule[0.8mm]
\end{tabular}}
\end{table}

Meanwhile, MoE improves both generation quality and diversity. When used in isolation, MoE slightly reduces FGD while significantly increasing diversity (Var.2). This indicates that it achieves richer gesture coverage without increasing feed-forward computation. Furthermore, MoE delivers consistent gains across different configurations. For instance, with keyframes but without interpolation, adding MoE improves both FGD and diversity (Var.6). Even when keyframes and the interpolator are already applied, MoE further lowers FGD (Var.7). This demonstrates that routed experts remain effective even when the sparse decoding mechanism is already in place.

\begin{table}[t]
\centering
\caption{Ablation study on keyframe stride for inference speed (N=5).}
\label{tab:exp_ablation_speed}
\scalebox{0.98}{
\begin{tabular}{ccc}
\midrule[0.8mm]
Keyframe Interval & FPS ($\uparrow$) & TTFF (ms) ($\downarrow$) \\
\midrule[0.5mm]
\midrule[0.5mm]
\multicolumn{3}{l}{\textit{Baseline}} \\
\midrule[0.5mm]
- & 19.14 & 6269 \\
\midrule[0.5mm]
\multicolumn{3}{l}{\textit{Linear Interpolation}} \\
\midrule[0.5mm]
s = 4 & 31.78 & 805 \\
s = 6 & 44.16 & 819 \\
s = 8 & 57.86 & 811 \\
\midrule[0.5mm]
\multicolumn{3}{l}{\textit{Network Interpolation}} \\
\midrule[0.5mm]
s = 4 & 31.54 & 811 \\
s = 6 & 43.82 & 826 \\
s = 6 (N=2) & 27.33 & 540 \\
s = 8 & 57.41 & 817 \\
\midrule[0.8mm]
\end{tabular}}
\end{table}

\begin{table}[t]
\centering
\caption{Runtime analysis for one forward pass of UMo.}
\label{tab:exp_runtime}
\scalebox{1}{
\begin{tabular}{lcc}
\midrule[0.8mm]
Stage & \multicolumn{2}{c}{Time (ms)} \\
\midrule[0.5mm]
- & N=5 & N=2 \\
\midrule[0.5mm]
\multicolumn{2}{l}{\textbf{\textit{Tokenizer}}} \\
\midrule[0.5mm]
Audio & 1.77 & 1.73 \\
Motion & 4.36 & 3.92 \\
\midrule[0.5mm]
\multicolumn{2}{l}{\textbf{\textit{Encoder}}} \\
\midrule[0.5mm]
Bidirectional Attention & 6.5 & 4.6 \\
MoE FFN & 19.0 & 17.9 \\
\midrule[0.5mm]
\multicolumn{2}{l}{\textbf{\textit{Decoder}}} \\
\midrule[0.5mm]
Causal Attention & 371.1 & 239.4 \\
MoE FFN & 232.3 & 189.9 \\
Other & 176.6 & 69.1 \\
Interpolation & 6.5 & 5.5 \\
\midrule[0.5mm]
\multicolumn{2}{l}{\textbf{\textit{Total}}} \\
\midrule[0.5mm]
- & 818 & 532\\
\midrule[0.8mm]
\end{tabular}}
\end{table}

Keyframe prediction without the learned interpolator outperforms the dense baseline (Var. 3), while pairing keyframes with the learned interpolator delivers the largest single-step drop in FGD (Var. 4). The same pattern holds when MoE is enabled (Var. 7). This confirms that predicting temporally sparse anchors reduces cumulative decoding error under a limited computational budget.

Incorporating all three modules, UMo achieves the best FGD (Var.9). Compared to Var8, the performance drop caused by the AR module is significantly mitigated, demonstrating the synergistic effect of UMo's modular design. Furthermore, introducing audio augmentation leads to a substantial increase in output diversity, with a slight decline in fitting metrics on the test set (Var.10). This aligns with our objective: audio augmentation is intended to guide the model in learning generalizable representations, rather than merely overfitting to specific audio-motion pairs. Simultaneously, shortening the decoding window length will bring time benefits, but it will lead to a decrease in output quality (Var.11).

\noindent\textbf{Training Recipe Analysis.} Table~\mref{tab:exp_ablation_recipe} presents the ablation study on the training recipe described in Sec.~\mref{sec:method_training}. Removing the multimodal pre-training stage yields the largest performance degradation, indicating that broad representation learning and auxiliary alignment signals are crucial for downstream co-speech quality. The results of removing post-training stage~1 suggest that dense instruction finetuning is also responsible for tightening audio-motion synchronization. In contrast, omitting the second stage of post-training has a marginal impact on these metrics. This is because its primary objective is to adapt the model to an autoregressive inference paradigm, the core value of which lies in significantly boosting inference speed and reducing response latency, rather than yielding substantial gains in generation quality.

\noindent\textbf{Keyframe Stride and Interpolation Analysis.}
Table~\mref{tab:exp_ablation_keyframe} compares linear interpolation with our learned interpolation network using strides $s\in\{4,6,8\}$. The quality of linear interpolation deteriorates as the stride grows. Specifically, FGD rises to 6.04 as $s{=}8$, reflecting the difficulty of interpolation task during sparse keyframe anchors. In contrast, our network reverses this trend. FGD improves to $5.13$ at $s{=}6$ and remains strong at $s{=}8$ ($5.12$). This indicates that sparse keyframes remain expressive when mid-interval motion is reconstructed by a lightweight network.

Table~\mref{tab:exp_ablation_speed} summarizes the inference speed differences, our keyframe settings is applied with chunk-wise autoregressive.
Compared to the dense decoding baseline, sparse keyframe schedules significantly reduce decoder workload. This reduces TTFF by roughly \textbf{$7.8\times$} and increases sustained throughput to \textbf{$31$--$58$} FPS depending on the stride $s$.
As $s$ increases from $4$ to $8$, FPS roughly doubles for both strategies.Specifically, it rises from $31.78$ to $57.86$ with linear interpolation, and from $31.54$ to $57.41$ with the network. This trend indicates that FPS is linearly correlated with the stride, suggesting that the primary time cost of UMo lies in the autoregressive per-token decoding step. However, the results for $N=2$ show that neither FPS nor TTFF is linearly correlated with the number of decoded frames per step, indicating that the speedup gains from compressing the prediction window size diminish gradually.
Replacing linear interpolation with the learned interpolation network adds minimal overhead. At each $s$, FPS differs by at most $0.7$ FPS and TTFF stays within $20$ ms.

\noindent\textbf{Run-Time Analysis.}
Table~\mref{tab:exp_runtime} lists wall-clock time per stage for one end-to-end forward pass of UMo on our profiling setup.
The full pass takes $818$ ms.
Tokenizer cost stays small: audio takes $1.77$ ms and motion takes $4.36$ ms.
The encoder adds $6.5$ ms for bidirectional attention and $19.0$ ms for encoder MoE FFNs.
In contrast, the decoder accounts for the bulk of latency: decoder attention takes $371.5$ ms, decoder MoE FFNs take $232.5$ ms, and remaining decoder-side paths take $176.6$ ms.
This allocation indicates that autoregressive decoding and decoder FFN blocks are the primary drivers of end-to-end latency, while tokenization and encoding are comparatively cheap.

The learned interpolation network adds only $6.5$ ms.
This matches Table~\mref{tab:exp_ablation_speed}, where linear versus network interpolation yields nearly identical FPS and TTFF at each stride $s$.
The results of $N=2$ suggest that the cost of UMo scales mainly with the number of decoder steps per forward and with decoder attention/MoE compute, not with filling intervals between keyframes.
Therefore, latency optimizations should prioritize sparse keyframe schedules, chunk horizons, and kernel-efficient decoder execution.

\textbf{\subsection{Limitations and Future Work}\label{sec:exp_limit}}

Although UMo has achieved significant progress in co-speech motion generation, several limitations remain to be addressed. First, the current temporal decoding strategy exhibits structural biases. The keyframe-centric design introduces a specific bias where prediction errors at anchor points can propagate through subsequent interpolation windows. Furthermore, the chunk-wise autoregressive generation couples output quality to the arbitrary boundaries of these windows, potentially affecting long-range temporal coherence. Additionally, there remains potential to further enhance the runtime efficiency of UMo, particularly in terms of inference speed and time to first frame.

Regarding the architectural choices, our current experiments have only scratched the surface in exploring the number of experts and activation counts. Future work will investigate a wider variety of combinations, while also incorporating expert parallelism to further boost inference speed, particularly under multi-expert activation settings.

Currently, our framework still relies on external audio inputs to drive the motion generation. As a direction for future research, we aim to explore audio generation capabilities, enabling the model to synthesize synchronized audio signals that match the generated motion style and rhythm, moving towards fully unified generation framework.

\section{\textbf{Conclusion}}

We present \textbf{UMo}, a unified autoregressive architecture that couples text, audio, and compositional motion tokens under unified optimization objectives. We employ chunk-wise autoregressive decoding with fixed context windows to optimize inference speed and response latency. To enhance performance, we introduce a MoE framework that provides spatial sparsity, increasing fidelity for body-part motion without incurring additional computational costs. Furthermore, our keyframe-centric generation paradigm introduces temporal sparsity to reduce the time redundancy caused by frame-wise dense predictions. Finally, a three-stage training recipe combined with targeted audio augmentation is applied to the entire architecture and ultimately achieved a balance between performance and efficiency. We explore the unification of paired multi-modal data and training objectives within an LLM backbone, ultimately achieving real-time motion generation. We hope our research will provide valuable insights and lay a solid foundation for future work.

\bibliography{arxiv}

\begin{thebibliography}{76}
\providecommand{\natexlab}[1]{#1}

\bibitem[{Alexanderson et~al.(2023)Alexanderson, Nagy, Beskow, and
  Henter}]{alexanderson2023listen}
Alexanderson, S.; Nagy, R.; Beskow, J.; and Henter, G.~E. 2023.
\newblock Listen, denoise, action! audio-driven motion synthesis with diffusion
  models.
\newblock \emph{ACM Transactions on Graphics (TOG)}, 42(4): 1--20.

\bibitem[{Ao et~al.(2022)Ao, Gao, Lou, Chen, and Liu}]{ao2022rhythmic}
Ao, T.; Gao, Q.; Lou, Y.; Chen, B.; and Liu, L. 2022.
\newblock Rhythmic gesticulator: Rhythm-aware co-speech gesture synthesis with
  hierarchical neural embeddings.
\newblock \emph{ACM Transactions on Graphics (TOG)}, 41(6): 1--19.

\bibitem[{Ao, Zhang, and Liu(2023)}]{ao2023gesturediffuclip}
Ao, T.; Zhang, Z.; and Liu, L. 2023.
\newblock Gesturediffuclip: Gesture diffusion model with clip latents.
\newblock \emph{ACM Transactions on Graphics (TOG)}, 42(4): 1--18.

\bibitem[{Bai et~al.(2023)Bai, Bai, Chu, Cui, Dang, Deng, Fan, Ge, Han, Huang
  et~al.}]{bai2023qwen}
Bai, J.; Bai, S.; Chu, Y.; Cui, Z.; Dang, K.; Deng, X.; Fan, Y.; Ge, W.; Han,
  Y.; Huang, F.; et~al. 2023.
\newblock Qwen technical report.
\newblock \emph{arXiv preprint arXiv:2309.16609}.

\bibitem[{Cai et~al.(2025)Cai, Chu, Gao, Gong, Huang, Kang, Li, Liu, Liu, Liu
  et~al.}]{cai2025mio}
Cai, Y.; Chu, X.; Gao, X.; Gong, S.; Huang, Y.; Kang, C.; Li, K.; Liu, H.; Liu,
  R.; Liu, Y.; et~al. 2025.
\newblock Towards Interactive Intelligence for Digital Humans.
\newblock \emph{arXiv preprint arXiv:2512.13674}.

\bibitem[{Chen et~al.(2024{\natexlab{a}})Chen, Li, Ding, Shao, and
  Zhou}]{chen2024enabling}
Chen, B.; Li, Y.; Ding, Y.-X.; Shao, T.; and Zhou, K. 2024{\natexlab{a}}.
\newblock Enabling synergistic full-body control in prompt-based co-speech
  motion generation.
\newblock In \emph{Proceedings of the 32nd ACM International Conference on
  Multimedia}, 6774--6783.

\bibitem[{Chen et~al.(2025)Chen, Zhang, Lakshmikanth, Fang, Shao, Wetzstein,
  Li, and Adeli}]{chen2024language}
Chen, C.; Zhang, J.; Lakshmikanth, S.~K.; Fang, Y.; Shao, R.; Wetzstein, G.;
  Li, F.-F.; and Adeli, E. 2025.
\newblock The Language of Motion: Unifying Verbal and Non-verbal Language of
  {3D} Human Motion.
\newblock In \emph{Proceedings of the {IEEE/CVF} Conference on Computer Vision
  and Pattern Recognition ({CVPR})}.

\bibitem[{Chen et~al.(2024{\natexlab{b}})Chen, Liu, Wang, Zeng, Li, and
  Chen}]{chen2024diffsheg}
Chen, J.; Liu, Y.; Wang, J.; Zeng, A.; Li, Y.; and Chen, Q. 2024{\natexlab{b}}.
\newblock Diffsheg: A diffusion-based approach for real-time speech-driven
  holistic 3d expression and gesture generation.
\newblock In \emph{Proceedings of the {IEEE/CVF} Conference on Computer Vision
  and Pattern Recognition ({CVPR})}, 7352--7361.

\bibitem[{Chen et~al.(2024{\natexlab{c}})Chen, Shi, Huang, Tan, Komura, and
  Chen}]{chen2024taming}
Chen, R.; Shi, M.; Huang, S.; Tan, P.; Komura, T.; and Chen, X.
  2024{\natexlab{c}}.
\newblock Taming diffusion probabilistic models for character control.
\newblock In \emph{ACM SIGGRAPH 2024 Conference Papers}, 1--10.

\bibitem[{Chen et~al.(2023)Chen, Jiang, Liu, Huang, Fu, Chen, and
  Yu}]{chen2024executing}
Chen, X.; Jiang, B.; Liu, W.; Huang, Z.; Fu, B.; Chen, T.; and Yu, G. 2023.
\newblock Executing your commands via motion diffusion in latent space.
\newblock In \emph{Proceedings of the {IEEE/CVF} Conference on Computer Vision
  and Pattern Recognition ({CVPR})}, 18000--18010.

\bibitem[{Chern et~al.(2025)Chern, Hu, Tang, Su, Chern, Deng, and
  Liu}]{chern2025livetalk}
Chern, E.; Hu, Z.; Tang, B.; Su, J.; Chern, S.; Deng, Z.; and Liu, P. 2025.
\newblock {LiveTalk}: Real-Time Multimodal Interactive Video Diffusion via
  Improved On-Policy Distillation.
\newblock \emph{arXiv preprint arXiv:2512.23576}.

\bibitem[{Chi et~al.(2024)Chi, Chi, Ma, Agarwal, Siddiqui, Ramani, and
  Lee}]{chi2024m2d2m}
Chi, S.; Chi, H.-g.; Ma, H.; Agarwal, N.; Siddiqui, F.; Ramani, K.; and Lee, K.
  2024.
\newblock M2d2m: Multi-motion generation from text with discrete diffusion
  models.
\newblock In \emph{Proceedings of the {European} Conference on Computer Vision
  ({ECCV})}, 18--36.

\bibitem[{D{\'e}fossez et~al.(2024)D{\'e}fossez, Mazar{\'e}, Orsini, Royer,
  P{\'e}rez, J{\'e}gou, Grave, and Zeghidour}]{defossez2024moshi}
D{\'e}fossez, A.; Mazar{\'e}, L.; Orsini, M.; Royer, A.; P{\'e}rez, P.;
  J{\'e}gou, H.; Grave, E.; and Zeghidour, N. 2024.
\newblock Moshi: a speech-text foundation model for real-time dialogue.
\newblock \emph{arXiv preprint arXiv:2410.00037}.

\bibitem[{Fu et~al.(2026)Fu, Sun, Fang, Cai, and Kim}]{fu2026mogo}
Fu, D.; Sun, T.; Fang, P.; Cai, X.; and Kim, H. 2026.
\newblock MOGO: Residual Quantized Hierarchical Causal Transformer for
  Real-Time and Infinite-Length 3D Human Motion Generation.
\newblock In \emph{Proceedings of the AAAI Conference on Artificial
  Intelligence (AAAI)}, 5, 3994--4002.

\bibitem[{Ginosar et~al.(2019)Ginosar, Bar, Kohavi, Chan, Owens, and
  Malik}]{ginosar2019learning}
Ginosar, S.; Bar, A.; Kohavi, G.; Chan, C.; Owens, A.; and Malik, J. 2019.
\newblock Learning individual styles of conversational gesture.
\newblock In \emph{Proceedings of the {IEEE/CVF} Conference on Computer Vision
  and Pattern Recognition ({CVPR})}, 3497--3506.

\bibitem[{Gu et~al.(2026)Gu, Zhang, Xie, Cai, Yang, and Liu}]{gu2026bridging}
Gu, C.; Zhang, M.; Xie, H.; Cai, Z.; Yang, L.; and Liu, Z. 2026.
\newblock Bridging Semantic and Kinematic Conditions with Diffusion-based
  Discrete Motion Tokenizer.
\newblock \emph{arXiv preprint arXiv:2603.19227}.

\bibitem[{Guo et~al.(2024)Guo, Mu, Javed, Wang, and Cheng}]{guo2024momask}
Guo, C.; Mu, Y.; Javed, M.~G.; Wang, S.; and Cheng, L. 2024.
\newblock Momask: Generative masked modeling of 3d human motions.
\newblock In \emph{Proceedings of the {IEEE/CVF} Conference on Computer Vision
  and Pattern Recognition ({CVPR})}, 1900--1910.

\bibitem[{Guo et~al.(2022{\natexlab{a}})Guo, Zou, Zuo, Wang, Ji, Li, and
  Cheng}]{guo2022humanml3d}
Guo, C.; Zou, S.; Zuo, X.; Wang, S.; Ji, W.; Li, X.; and Cheng, L.
  2022{\natexlab{a}}.
\newblock Generating diverse and natural 3d human motions from text.
\newblock In \emph{Proceedings of the {IEEE/CVF} Conference on Computer Vision
  and Pattern Recognition ({CVPR})}, 5152--5161.

\bibitem[{Guo et~al.(2022{\natexlab{b}})Guo, Zuo, Wang, and
  Cheng}]{guo2022tm2t}
Guo, C.; Zuo, X.; Wang, S.; and Cheng, L. 2022{\natexlab{b}}.
\newblock Tm2t: Stochastic and tokenized modeling for the reciprocal generation
  of 3d human motions and texts.
\newblock In \emph{Proceedings of the {European} Conference on Computer Vision
  ({ECCV})}, 580--597.

\bibitem[{Habibie et~al.(2021)Habibie, Xu, Mehta, Liu, Seidel, Pons-Moll,
  Elgharib, and Theobalt}]{habibie2021learning}
Habibie, I.; Xu, W.; Mehta, D.; Liu, L.; Seidel, H.-P.; Pons-Moll, G.;
  Elgharib, M.; and Theobalt, C. 2021.
\newblock Learning speech-driven 3d conversational gestures from video.
\newblock In \emph{Proceedings of the 21st ACM international conference on
  intelligent virtual agents}, 101--108.

\bibitem[{Han et~al.(2025)Han, Teye, Yadgaroff, and B{\"u}tepage}]{han2025tiny}
Han, Z.; Teye, M.; Yadgaroff, D.; and B{\"u}tepage, J. 2025.
\newblock Tiny is not small enough: High quality, low-resource facial animation
  models through hybrid knowledge distillation.
\newblock \emph{ACM Transactions on Graphics (TOG)}, 44(4): 1--18.

\bibitem[{Hou et~al.(2025)Hou, Luo, Pan, Chang, and Shan}]{hou2025motionverse}
Hou, R.; Luo, M.; Pan, H.; Chang, H.; and Shan, S. 2025.
\newblock Motionverse: A unified multimodal framework for motion comprehension,
  generation and editing.
\newblock \emph{arXiv preprint arXiv:2509.23635}.

\bibitem[{Hsu et~al.(2021)Hsu, Bolte, Tsai, Lakhotia, Salakhutdinov, and
  Mohamed}]{hsu2021hubert}
Hsu, W.-N.; Bolte, B.; Tsai, Y.-H.~H.; Lakhotia, K.; Salakhutdinov, R.; and
  Mohamed, A. 2021.
\newblock Hubert: Self-supervised speech representation learning by masked
  prediction of hidden units.
\newblock \emph{IEEE/ACM transactions on audio, speech, and language
  processing}, 29: 3451--3460.

\bibitem[{Jiang et~al.(2023)Jiang, Chen, Liu, Yu, Yu, and
  Chen}]{jiang2024motiongpt}
Jiang, B.; Chen, X.; Liu, W.; Yu, J.; Yu, G.; and Chen, T. 2023.
\newblock Motiongpt: Human motion as a foreign language.
\newblock In \emph{Advances in Neural Information Processing Systems
  ({NeurIPS})}, volume~36, 20067--20079.

\bibitem[{Karunratanakul et~al.(2023)Karunratanakul, Preechakul, Suwajanakorn,
  and Tang}]{karunratanakul2023guided}
Karunratanakul, K.; Preechakul, K.; Suwajanakorn, S.; and Tang, S. 2023.
\newblock Guided motion diffusion for controllable human motion synthesis.
\newblock In \emph{Proceedings of the {IEEE/CVF} International Conference on
  Computer Vision ({ICCV})}, 2151--2162.

\bibitem[{Li et~al.(2021)Li, Yang, Ross, and Kanazawa}]{li2021ai}
Li, R.; Yang, S.; Ross, D.~A.; and Kanazawa, A. 2021.
\newblock Ai choreographer: Music conditioned 3d dance generation with aist++.
\newblock In \emph{Proceedings of the {IEEE/CVF} Conference on Computer Vision
  and Pattern Recognition ({CVPR})}, 13401--13412.

\bibitem[{Li et~al.(2026)Li, An, Tang, Guo, Shugurov, Zhang, Zhao, Sridhar,
  Tao, and Mittal}]{li2026llamo}
Li, Z.; An, S.; Tang, C.; Guo, C.; Shugurov, I.; Zhang, L.; Zhao, A.; Sridhar,
  S.; Tao, L.; and Mittal, A. 2026.
\newblock LLaMo: Scaling Pretrained Language Models for Unified Motion
  Understanding and Generation with Continuous Autoregressive Tokens.
\newblock In \emph{Proceedings of the {IEEE/CVF} Conference on Computer Vision
  and Pattern Recognition ({CVPR})}.

\bibitem[{Lin et~al.(2023)Lin, Zeng, Lu, Cai, Zhang, Wang, and
  Zhang}]{lin2023motionx}
Lin, J.; Zeng, A.; Lu, S.; Cai, Y.; Zhang, R.; Wang, H.; and Zhang, L. 2023.
\newblock Motion-x: A large-scale 3d expressive whole-body human motion
  dataset.
\newblock \emph{Advances in Neural Information Processing Systems}, 36:
  25268--25280.

\bibitem[{Ling et~al.(2024)Ling, Han, Li, Cheng, Shen, and
  Zou}]{ling2024versatilemotion}
Ling, Z.; Han, B.; Li, S.; Cheng, J.; Shen, H.; and Zou, C. 2024.
\newblock VersatileMotion: A Unified Framework for Motion Synthesis and
  Comprehension.
\newblock \emph{arXiv preprint arXiv:2411.17335}.

\bibitem[{Liu et~al.(2022{\natexlab{a}})Liu, Iwamoto, Zhu, Li, Zhou, Bozkurt,
  and Zheng}]{liu2022disco}
Liu, H.; Iwamoto, N.; Zhu, Z.; Li, Z.; Zhou, Y.; Bozkurt, E.; and Zheng, B.
  2022{\natexlab{a}}.
\newblock Disco: Disentangled implicit content and rhythm learning for diverse
  co-speech gestures synthesis.
\newblock In \emph{Proceedings of the 30th ACM international conference on
  multimedia}, 3764--3773.

\bibitem[{Liu et~al.(2024)Liu, Zhu, Becherini, Peng, Su, Zhou, Zhe, Iwamoto,
  Zheng, and Black}]{liu2024emage}
Liu, H.; Zhu, Z.; Becherini, G.; Peng, Y.; Su, M.; Zhou, Y.; Zhe, X.; Iwamoto,
  N.; Zheng, B.; and Black, M.~J. 2024.
\newblock {EMAGE}: Towards Unified Holistic Co-Speech Gesture Generation via
  Expressive Masked Audio Gesture Modeling.
\newblock In \emph{Proceedings of the {IEEE/CVF} Conference on Computer Vision
  and Pattern Recognition ({CVPR})}, 1144--1154.

\bibitem[{Liu et~al.(2022{\natexlab{b}})Liu, Zhu, Iwamoto, Peng, Li, Zhou,
  Bozkurt, and Zheng}]{liu2022beat}
Liu, H.; Zhu, Z.; Iwamoto, N.; Peng, Y.; Li, Z.; Zhou, Y.; Bozkurt, E.; and
  Zheng, B. 2022{\natexlab{b}}.
\newblock Beat: A large-scale semantic and emotional multi-modal dataset for
  conversational gestures synthesis.
\newblock In \emph{Proceedings of the {European} Conference on Computer Vision
  ({ECCV})}, 612--630.

\bibitem[{Liu et~al.(2025)Liu, Song, Huang, and Xu}]{liu2025gesturelsm}
Liu, P.; Song, L.; Huang, J.; and Xu, C. 2025.
\newblock {GestureLSM}: Latent Shortcut based Co-Speech Gesture Generation with
  Spatial-Temporal Modeling.
\newblock In \emph{Proceedings of the {IEEE/CVF} International Conference on
  Computer Vision ({ICCV})}.

\bibitem[{Liu et~al.(2022{\natexlab{c}})Liu, Wu, Zhou, Du, Wu, Lin, and
  Liu}]{liu2022audio}
Liu, X.; Wu, Q.; Zhou, H.; Du, Y.; Wu, W.; Lin, D.; and Liu, Z.
  2022{\natexlab{c}}.
\newblock Audio-driven co-speech gesture video generation.
\newblock In \emph{Advances in Neural Information Processing Systems
  ({NeurIPS})}, volume~35, 21386--21399.

\bibitem[{Liu et~al.(2022{\natexlab{d}})Liu, Wu, Zhou, Xu, Qian, Lin, Zhou, Wu,
  Dai, and Zhou}]{liu2022learning}
Liu, X.; Wu, Q.; Zhou, H.; Xu, Y.; Qian, R.; Lin, X.; Zhou, X.; Wu, W.; Dai,
  B.; and Zhou, B. 2022{\natexlab{d}}.
\newblock Learning hierarchical cross-modal association for co-speech gesture
  generation.
\newblock In \emph{Proceedings of the {IEEE/CVF} Conference on Computer Vision
  and Pattern Recognition ({CVPR})}, 10462--10472.

\bibitem[{Lu et~al.(2025)Lu, Wang, Lu, Chen, Dai, Dong, Dou, Dai, and
  Zhang}]{lu2025scamo}
Lu, S.; Wang, J.; Lu, Z.; Chen, L.-H.; Dai, W.; Dong, J.; Dou, Z.; Dai, B.; and
  Zhang, R. 2025.
\newblock Scamo: Exploring the scaling law in autoregressive motion generation
  model.
\newblock In \emph{Proceedings of the {IEEE/CVF} Conference on Computer Vision
  and Pattern Recognition ({CVPR})}, 27872--27882.

\bibitem[{Mahmood et~al.(2019)Mahmood, Ghorbani, Troje, Pons-Moll, and
  Black}]{mahmood2019amass}
Mahmood, N.; Ghorbani, N.; Troje, N.~F.; Pons-Moll, G.; and Black, M.~J. 2019.
\newblock AMASS: Archive of motion capture as surface shapes.
\newblock In \emph{Proceedings of the IEEE/CVF International Conference on
  Computer Vision (ICCV)}, 5442--5451.

\bibitem[{{MiniMax}(2025)}]{minimax_speech_2025}
{MiniMax}. 2025.
\newblock MiniMax Speech 2.6.

\bibitem[{Mughal et~al.(2026)Mughal, Dabral, Demberg, and
  Theobalt}]{mughal2026miburi}
Mughal, M.~H.; Dabral, R.; Demberg, V.; and Theobalt, C. 2026.
\newblock {MIBURI}: Towards Expressive Interactive Gesture Synthesis.
\newblock In \emph{Proceedings of the {IEEE/CVF} Conference on Computer Vision
  and Pattern Recognition ({CVPR})}.

\bibitem[{Mughal et~al.(2025)Mughal, Dabral, Scholman, Demberg, and
  Theobalt}]{mughal2025retrieving}
Mughal, M.~H.; Dabral, R.; Scholman, M.~C.; Demberg, V.; and Theobalt, C. 2025.
\newblock Retrieving semantics from the deep: an rag solution for gesture
  synthesis.
\newblock In \emph{Proceedings of the {IEEE/CVF} Conference on Computer Vision
  and Pattern Recognition ({CVPR})}, 16578--16588.

\bibitem[{Nazarenus et~al.(2026)Nazarenus, Li, He, Xie, Lenssen, and
  Pons-Moll}]{nazarenus2026actionplan}
Nazarenus, E.; Li, C.; He, Y.; Xie, X.; Lenssen, J.~E.; and Pons-Moll, G. 2026.
\newblock ActionPlan: Future-Aware Streaming Motion Synthesis via Frame-Level
  Action Planning.
\newblock \emph{arXiv preprint arXiv:2603.13500}.

\bibitem[{Pan, Singh, and Hafemann(2025)}]{pan2025model}
Pan, Y.; Singh, K.; and Hafemann, L.~G. 2025.
\newblock Model See Model Do: Speech-Driven Facial Animation with Style
  Control.
\newblock In \emph{Proceedings of the Special Interest Group on Computer
  Graphics and Interactive Techniques Conference Conference Papers}, 1--10.

\bibitem[{Petrovich, Black, and Varol(2022)}]{petrovich2021temos}
Petrovich, M.; Black, M.~J.; and Varol, G. 2022.
\newblock Temos: Generating diverse human motions from textual descriptions.
\newblock In \emph{Proceedings of the {European} Conference on Computer Vision
  ({ECCV})}, 480--497.

\bibitem[{Pinyoanuntapong et~al.(2024)Pinyoanuntapong, Wang, Lee, and
  Chen}]{pinyoanuntapong2024mmm}
Pinyoanuntapong, E.; Wang, P.; Lee, M.; and Chen, C. 2024.
\newblock Mmm: Generative masked motion model.
\newblock In \emph{Proceedings of the {IEEE/CVF} Conference on Computer Vision
  and Pattern Recognition ({CVPR})}, 1546--1555.

\bibitem[{Punnakkal et~al.(2021)Punnakkal, Chandrasekaran, Athanasiou,
  Quiros-Ramirez, and Black}]{punnakkal2021babel}
Punnakkal, A.~R.; Chandrasekaran, A.; Athanasiou, N.; Quiros-Ramirez, A.; and
  Black, M.~J. 2021.
\newblock BABEL: Bodies, action and behavior with english labels.
\newblock In \emph{Proceedings of the {IEEE/CVF} Conference on Computer Vision
  and Pattern Recognition ({CVPR})}, 722--731.

\bibitem[{Raffel et~al.(2020)Raffel, Shazeer, Roberts, Lee, Narang, Matena,
  Zhou, Li, and Liu}]{raffel2020exploring}
Raffel, C.; Shazeer, N.; Roberts, A.; Lee, K.; Narang, S.; Matena, M.; Zhou,
  Y.; Li, W.; and Liu, P.~J. 2020.
\newblock Exploring the limits of transfer learning with a unified text-to-text
  transformer.
\newblock \emph{Journal of machine learning research}, 21(140): 1--67.

\bibitem[{Rempe et~al.(2026)Rempe, Petrovich, Yuan, Zhang, Peng, Jiang, Wang,
  Iqbal, Minor, de~Ruyter et~al.}]{rempe2026kimodo}
Rempe, D.; Petrovich, M.; Yuan, Y.; Zhang, H.; Peng, X.~B.; Jiang, Y.; Wang,
  T.; Iqbal, U.; Minor, D.; de~Ruyter, M.; et~al. 2026.
\newblock Kimodo: Scaling Controllable Human Motion Generation.
\newblock \emph{arXiv preprint arXiv:2603.15546}.

\bibitem[{Sun et~al.(2024)Sun, Lv, Ye, Lin, Sheng, Wen, Yu, and
  Liu}]{sun2024diffposetalk}
Sun, Z.; Lv, T.; Ye, S.; Lin, M.; Sheng, J.; Wen, Y.-H.; Yu, M.; and Liu, Y.-j.
  2024.
\newblock Diffposetalk: Speech-driven stylistic 3d facial animation and head
  pose generation via diffusion models.
\newblock \emph{ACM Transactions on Graphics (ToG)}, 43(4): 1--9.

\bibitem[{Tevet et~al.(2022)Tevet, Gordon, Hertz, Bermano, and
  Cohen-Or}]{tevet2022motionclip}
Tevet, G.; Gordon, B.; Hertz, A.; Bermano, A.~H.; and Cohen-Or, D. 2022.
\newblock Motionclip: Exposing human motion generation to clip space.
\newblock In \emph{Proceedings of the {European} Conference on Computer Vision
  ({ECCV})}, 358--374.

\bibitem[{Tevet et~al.(2024)Tevet, Raab, Cohan, Reda, Luo, Peng, Bermano, and
  van~de Panne}]{tevet2024closd}
Tevet, G.; Raab, S.; Cohan, S.; Reda, D.; Luo, Z.; Peng, X.~B.; Bermano, A.~H.;
  and van~de Panne, M. 2024.
\newblock Closd: Closing the loop between simulation and diffusion for
  multi-task character control.
\newblock \emph{arXiv preprint arXiv:2410.03441}.

\bibitem[{Tevet et~al.(2023)Tevet, Raab, Gordon, Shafir, Cohen-Or, and
  Bermano}]{tevet2023human}
Tevet, G.; Raab, S.; Gordon, B.; Shafir, Y.; Cohen-Or, D.; and Bermano, A.~H.
  2023.
\newblock Human Motion Diffusion Model.
\newblock In \emph{International Conference on Learning Representations
  ({ICLR})}.

\bibitem[{Touvron et~al.(2023)Touvron, Lavril, Izacard, Martinet, Lachaux,
  Lacroix, Rozi{\`e}re, Goyal, Hambro, Azhar et~al.}]{touvron2023llama}
Touvron, H.; Lavril, T.; Izacard, G.; Martinet, X.; Lachaux, M.-A.; Lacroix,
  T.; Rozi{\`e}re, B.; Goyal, N.; Hambro, E.; Azhar, F.; et~al. 2023.
\newblock Llama: Open and efficient foundation language models.
\newblock \emph{arXiv preprint arXiv:2302.13971}.

\bibitem[{Wu et~al.(2025)Wu, Zhao, Wang, Liu, Tai, and
  Tang}]{wu2024motionagent}
Wu, Q.; Zhao, Y.; Wang, Y.; Liu, X.; Tai, Y.-W.; and Tang, C.-K. 2025.
\newblock {Motion-Agent}: A Conversational Framework for Human Motion
  Generation with {LLM}s.
\newblock In \emph{International Conference on Learning Representations
  ({ICLR})}.

\bibitem[{Xiao et~al.(2025)Xiao, Lu, Pi, Fan, Pan, Zhou, Feng, Zhou, Peng, and
  Wang}]{xiao2025motionstreamer}
Xiao, L.; Lu, S.; Pi, H.; Fan, K.; Pan, L.; Zhou, Y.; Feng, Z.; Zhou, X.; Peng,
  S.; and Wang, J. 2025.
\newblock Motionstreamer: Streaming motion generation via diffusion-based
  autoregressive model in causal latent space.
\newblock In \emph{Proceedings of the IEEE/CVF International Conference on
  Computer Vision (ICCV)}, 10086--10096.

\bibitem[{Xie et~al.(2024)Xie, Jampani, Zhong, Sun, and
  Jiang}]{xie2024omnicontrol}
Xie, Y.; Jampani, V.; Zhong, L.; Sun, D.; and Jiang, H. 2024.
\newblock Omnicontrol: Control any joint at any time for human motion
  generation.
\newblock In \emph{International Conference on Learning Representations
  ({ICLR})}.

\bibitem[{Xing et~al.(2023)Xing, Xia, Zhang, Cun, Wang, and
  Wong}]{xing2023codetalker}
Xing, J.; Xia, M.; Zhang, Y.; Cun, X.; Wang, J.; and Wong, T.-T. 2023.
\newblock Codetalker: Speech-driven 3d facial animation with discrete motion
  prior.
\newblock In \emph{Proceedings of the {IEEE/CVF} Conference on Computer Vision
  and Pattern Recognition ({CVPR})}, 12780--12790.

\bibitem[{Xu et~al.(2024)Xu, Lin, Han, Yang, Li, Zhang, and
  Li}]{xu2024mambatalk}
Xu, Z.; Lin, Y.; Han, H.; Yang, S.; Li, R.; Zhang, Y.; and Li, X. 2024.
\newblock Mambatalk: Efficient holistic gesture synthesis with selective state
  space models.
\newblock In \emph{Advances in Neural Information Processing Systems
  ({NeurIPS})}, volume~37, 20055--20080.

\bibitem[{Yi et~al.(2023)Yi, Liang, Liu, Cao, Wen, Bolkart, Tao, and
  Black}]{yi2023generating}
Yi, H.; Liang, H.; Liu, Y.; Cao, Q.; Wen, Y.; Bolkart, T.; Tao, D.; and Black,
  M.~J. 2023.
\newblock Generating holistic 3d human motion from speech.
\newblock In \emph{Proceedings of the {IEEE/CVF} Conference on Computer Vision
  and Pattern Recognition ({CVPR})}, 469--480.

\bibitem[{Yu et~al.(2025)Yu, Zhang, Chen, Xiang, Fang, Niebles, and
  Adeli}]{yu2025socialgen}
Yu, H.; Zhang, J.; Chen, C.; Xiang, T.; Fang, Y.; Niebles, J.~C.; and Adeli, E.
  2025.
\newblock Socialgen: Modeling multi-human social interaction with language
  models.
\newblock \emph{arXiv preprint arXiv:2503.22906}.

\bibitem[{Yuan et~al.(2023)Yuan, Song, Iqbal, Vahdat, and
  Kautz}]{yuan2023physdiff}
Yuan, Y.; Song, J.; Iqbal, U.; Vahdat, A.; and Kautz, J. 2023.
\newblock Physdiff: Physics-guided human motion diffusion model.
\newblock In \emph{Proceedings of the {IEEE/CVF} International Conference on
  Computer Vision ({ICCV})}, 16010--16021.

\bibitem[{Zeng et~al.(2025)Zeng, Huang, Wu, and Zheng}]{zeng2025light}
Zeng, L.-A.; Huang, G.; Wu, G.; and Zheng, W.-S. 2025.
\newblock Light-t2m: A lightweight and fast model for text-to-motion
  generation.
\newblock In \emph{Proceedings of the AAAI Conference on Artificial
  Intelligence (AAAI)}, 9, 9797--9805.

\bibitem[{Zhang, Fan, and Yang(2025)}]{zhang2025energymogen}
Zhang, J.; Fan, H.; and Yang, Y. 2025.
\newblock Energymogen: Compositional human motion generation with energy-based
  diffusion model in latent space.
\newblock In \emph{Proceedings of the {IEEE/CVF} Conference on Computer Vision
  and Pattern Recognition ({CVPR})}, 17592--17602.

\bibitem[{Zhang et~al.(2023{\natexlab{a}})Zhang, Zhang, Cun, Zhang, Zhao, Lu,
  Shen, and Shan}]{zhang2023generating}
Zhang, J.; Zhang, Y.; Cun, X.; Zhang, Y.; Zhao, H.; Lu, H.; Shen, X.; and Shan,
  Y. 2023{\natexlab{a}}.
\newblock Generating human motion from textual descriptions with discrete
  representations.
\newblock In \emph{Proceedings of the {IEEE/CVF} Conference on Computer Vision
  and Pattern Recognition ({CVPR})}, 14730--14740.

\bibitem[{Zhang et~al.(2024{\natexlab{a}})Zhang, Cai, Pan, Hong, Guo, Yang, and
  Liu}]{zhang2022motiondiffuse}
Zhang, M.; Cai, Z.; Pan, L.; Hong, F.; Guo, X.; Yang, L.; and Liu, Z.
  2024{\natexlab{a}}.
\newblock Motiondiffuse: Text-driven human motion generation with diffusion
  model.
\newblock \emph{IEEE Transactions on Pattern Analysis and Machine
  Intelligence}, 46(6): 4115--4128.

\bibitem[{Zhang et~al.(2023{\natexlab{b}})Zhang, Guo, Pan, Cai, Hong, Li, Yang,
  and Liu}]{zhang2023remodiffuse}
Zhang, M.; Guo, X.; Pan, L.; Cai, Z.; Hong, F.; Li, H.; Yang, L.; and Liu, Z.
  2023{\natexlab{b}}.
\newblock Remodiffuse: Retrieval-augmented motion diffusion model.
\newblock In \emph{Proceedings of the {IEEE/CVF} International Conference on
  Computer Vision ({ICCV})}, 364--373.

\bibitem[{Zhang et~al.(2023{\natexlab{c}})Zhang, Li, Cai, Ren, Yang, and
  Liu}]{zhang2023finemogen}
Zhang, M.; Li, H.; Cai, Z.; Ren, J.; Yang, L.; and Liu, Z. 2023{\natexlab{c}}.
\newblock Finemogen: Fine-grained spatio-temporal motion generation and
  editing.
\newblock In \emph{Advances in Neural Information Processing Systems
  ({NeurIPS})}, volume~36, 13981--13992.

\bibitem[{Zhang et~al.(2026)Zhang, Li, Loh, Xu, Wang, Wen, He, Zhao, Gong, and
  Zhang}]{zhang2026dimo}
Zhang, N.; Li, Z.; Loh, K.~W.; Xu, M.; Wang, Q.; Wen, Z.; He, X.; Zhao, W.;
  Gong, K.; and Zhang, M. 2026.
\newblock DiMo: Discrete Diffusion Modeling for Motion Generation and
  Understanding.
\newblock \emph{arXiv preprint arXiv:2602.04188}.

\bibitem[{Zhang et~al.(2025{\natexlab{a}})Zhang, Feng, Cseke, Saini, Bajandas,
  Heron, and Black}]{zhang2025primal}
Zhang, Y.; Feng, Y.; Cseke, A.; Saini, N.; Bajandas, N.; Heron, N.; and Black,
  M.~J. 2025{\natexlab{a}}.
\newblock {PRIMAL}: Physically Reactive and Interactive Motor Model for Avatar
  Learning.
\newblock In \emph{Proceedings of the {IEEE/CVF} International Conference on
  Computer Vision ({ICCV})}, 12725--12736.

\bibitem[{Zhang et~al.(2024{\natexlab{b}})Zhang, Huang, Liu, Tang, Lu, Chen,
  Bai, Chu, Yu, and Ouyang}]{zhang2024motiongptfinetuned}
Zhang, Y.; Huang, D.; Liu, B.; Tang, S.; Lu, Y.; Chen, L.; Bai, L.; Chu, Q.;
  Yu, N.; and Ouyang, W. 2024{\natexlab{b}}.
\newblock {MotionGPT}: Finetuned {LLM}s Are General-Purpose Motion Generators.
\newblock In \emph{Proceedings of the {AAAI} Conference on Artificial
  Intelligence (AAAI)}, 7, 7368--7376.

\bibitem[{Zhang et~al.(2025{\natexlab{b}})Zhang, Sun, Fang, Wang, Cai, Zhang,
  and Kim}]{zhang2025motionduet}
Zhang, Y.-Y.; Sun, T.; Fang, P.; Wang, D.-B.; Cai, X.; Zhang, M.-L.; and Kim,
  H. 2025{\natexlab{b}}.
\newblock MotionDuet: Dual-Conditioned 3D Human Motion Generation with
  Video-Regularized Text Learning.
\newblock \emph{arXiv preprint arXiv:2511.18209}.

\bibitem[{Zhang et~al.(2024{\natexlab{c}})Zhang, Ao, Zhang, Gao, Lin, Chen, and
  Liu}]{zhang2024semantic}
Zhang, Z.; Ao, T.; Zhang, Y.; Gao, Q.; Lin, C.; Chen, B.; and Liu, L.
  2024{\natexlab{c}}.
\newblock Semantic gesticulator: Semantics-aware co-speech gesture synthesis.
\newblock \emph{ACM Transactions on Graphics (TOG)}, 43(4): 1--17.

\bibitem[{Zhao, Li, and Tang(2025)}]{zhao2025dartcontrol}
Zhao, K.; Li, G.; and Tang, S. 2025.
\newblock DartControl: A Diffusion-Based Autoregressive Motion Model for
  Real-Time Text-Driven Motion Control.
\newblock In \emph{International Conference on Learning Representations
  ({ICLR})}.

\bibitem[{Zhao et~al.(2024)Zhao, Long, Zhang, Qin, Liang, Zhang, Zhang, Yu, and
  Xu}]{zhao2024media2face}
Zhao, Q.; Long, P.; Zhang, Q.; Qin, D.; Liang, H.; Zhang, L.; Zhang, Y.; Yu,
  J.; and Xu, L. 2024.
\newblock Media2face: Co-speech facial animation generation with multi-modality
  guidance.
\newblock In \emph{ACM SIGGRAPH 2024 conference papers}, 1--13.

\bibitem[{Zheng et~al.(2025)Zheng, Chen, Yao, Zeng, Jiang, Wang, Lasenby, and
  Jin}]{zheng2025autokeyframe}
Zheng, B.; Chen, K.; Yao, Y.; Zeng, Z.; Jiang, X.; Wang, H.; Lasenby, J.; and
  Jin, X. 2025.
\newblock Autokeyframe: Autoregressive keyframe generation for human motion
  synthesis and editing.
\newblock In \emph{Proceedings of the Special Interest Group on Computer
  Graphics and Interactive Techniques Conference Conference Papers}, 1--12.

\bibitem[{Zhou et~al.(2024)Zhou, Dou, Cao, Liao, Wang, Wang, Liu, Komura, Wang,
  and Liu}]{zhou2024emdm}
Zhou, W.; Dou, Z.; Cao, Z.; Liao, Z.; Wang, J.; Wang, W.; Liu, Y.; Komura, T.;
  Wang, W.; and Liu, L. 2024.
\newblock Emdm: Efficient motion diffusion model for fast and high-quality
  motion generation.
\newblock In \emph{Proceedings of the {European} Conference on Computer Vision
  ({ECCV})}, 18--38.

\bibitem[{Zhu et~al.(2025)Zhu, Jiang, Wang, Tang, Chen, Luo, Zheng, and
  Chen}]{zhu2025motiongpt3}
Zhu, B.; Jiang, B.; Wang, S.; Tang, S.; Chen, T.; Luo, L.; Zheng, Y.; and Chen,
  X. 2025.
\newblock Motiongpt3: Human motion as a second modality.
\newblock \emph{arXiv preprint arXiv:2506.24086}.

\end{thebibliography}

\end{document}